\documentclass[prx,twocolumn, amsfonts,amsmath, floatfix]{revtex4}
\usepackage{graphicx}
\usepackage{epsfig}
\usepackage{epstopdf}
\usepackage[usenames,dvipsnames]{xcolor}
\usepackage{braket}
\usepackage{hyperref}
\usepackage{amsthm}
\usepackage{enumitem}

\newcommand{\be}{\begin{equation}}
\newcommand{\ee}{\end{equation}}
\newcommand{\ba}{\begin{aligned}}
\newcommand{\ea}{\end{aligned}}

\newcommand{\bc}{\begin{center}}
\newcommand{\ec}{\end{center}}
\newcommand{\beq}{\begin{equation}}
\newcommand{\eeq}{\end{equation}}
\newcommand{\beqq}{\begin{equation*}}
\newcommand{\eeqq}{\end{equation*}}
\newcommand{\beqa}{\begin{align}}
\newcommand{\eeqa}{\end{align}}
\newcommand{\barr}{\begin{array}}
\newcommand{\earr}{\end{array}}
\newcommand{\bi}{\begin{itemize}}
\newcommand{\ei}{\end{itemize}}

\newcommand{\e}{\ensuremath{\mathrm{e}}}

\begin{document}

\title{Optimal quantum-programmable projective measurements with coherent states}

\author{Niraj Kumar$^{1}$}
\email{nkumar@exseed.ed.ac.uk}
\author{Ulysse Chabaud$^2$}
\email{ulysse.chabaud@gmail.com}
\author{Elham Kashefi$^{1,2}$}
\email{ekashefi@exseed.ed.ac.uk}
\author{Damian Markham$^{2,3}$}
\email{damian.markham@lip6.fr}
\author{Eleni Diamanti$^2$}
\email{eleni.diamanti@lip6.fr}
\address{$^1$School of Informatics, 10 Crichton St., University of Edinburgh, United Kingdom, \\ $^2$CNRS, LIP6, Sorbonne Universit\'{e}, 4 place Jussieu, 75005 Paris, France, \\ $^3$JFLI, CNRS, National Institute of Informatics, University of Tokyo, Tokyo, Japan}
\date{\today}


\begin{abstract}

We consider a device which can be programmed using coherent states of light to approximate a given projective measurement on an input coherent state. We provide and discuss three practical implementations of this programmable projective measurement device with linear optics, involving only balanced beam splitters and single photon threshold detectors. The three schemes optimally approximate any projective measurement onto a program coherent state in a non-destructive fashion. 
We further extend these to the case where there are no assumptions on the input state. In this setting, we show that our scheme enables an efficient verification of an unbounded untrusted source with only local coherent states, balanced beam splitters, and threshold detectors. Exploiting the link between programmable measurements and generalised swap test, we show as a direct application that our schemes provide an asymptotically quadratic improvement in existing quantum fingerprinting protocol to approximate the Euclidean distance between two unit vectors.

\end{abstract}

\maketitle


\section{Introduction}

\noindent A typical experiment with measurement of a quantum state involves preparing a circuit, which is often classically controlled, to perform a vast range of operations on the state. 
The choice of measurement typically depends on the task at hand and is usually fixed beforehand. In some cases, one could alternatively use a measurement device which is classically (re)programmable to obtain a wide variety of measurement scenarios. Recent developments on this front involve a programmable optical circuit that can implement all possible linear optical protocols up to the size of that circuit~\cite{carolan2015universal}. The circuit involves Mach--Zehnder interferometers and thermo-optic phase shifters which are electronically and optically controlled. \par

In this work, we investigate an alternative scenario where a quantum input---rather than a classical program---controls the choice of measurement. This setting has applications in several quantum protocols for solving communication complexity problems. These include quantum fingerprinting protocols to check for equality between two given strings ~\cite{buhrman2001quantum, arrazola2014quantum, guan2016observation, guan2018experimental}, Euclidean distance of two real vectors ~\cite{kumar2017efficient}, and matching-based one-way communication complexity problems~\cite{bar2004exponential,kumar2019experimental}. The choice of measurement being driven by an input quantum state has also been extensively used in cryptography settings such as private quantum money schemes~\cite{georgiou2015new, amiri2017quantum, guan2018experimental, bozzio2018experimental, kumar2019practically}. In these works, a central issue is to test whether two unknown quantum states are equal. Such a quantum state comparison can be performed using a programmable projective measurement device, where multiple copies of one of the two states to be compared act as program states encoding the direction of the measurement, which is then performed on the other state: if the projection succeeds, the states are considered equal, otherwise they are considered different.\par

The comparison of two states is trivial in the classical world, where two bit strings can be bit-wise checked and thus the maximum number of operations that are needed equals the size of the strings. However, comparing two quantum states is non-trivial since a quantum state is typically in a superposition over multiple possible basis states. If we simply follow the above classical procedure of individually measuring the quantum states in some fixed basis and comparing the measurement outcomes,  then the basis measurement only reveals partial information about each quantum state, i.e., the amplitude corresponding to the specific basis state onto which the quantum state has collapsed. Hence, this measure-and-compare approach does not work for comparing two unknown quantum states. Buhrman et al.~\cite{buhrman2001quantum} introduced a simple test, the so-called swap test, to compare two unknown quantum states with one-sided error probability, i.e., the test succeeds with certainty if the two quantum states are the same, however, there is a nonzero probability of failing the test if the two states are different.  This test uses a controlled-swap operation and is optimal under one-sided error probability, if one only has a single copy of the two quantum states~\cite{barnett2003comparison}. However, to succeed with an arbitrarily small desired error probability $\epsilon$, this technique needs to perform independent tests on at-least logarithm of inverse-$\epsilon$ number of copies of both quantum states. \par
 
Generalising this scenario, Chabaud et al.~\cite{chabaud2018optimal} introduced a version of the swap test when one is provided with just a single copy of one of the states in some \emph{input} register, and multiple copies of the other state in the \emph{program} registers. They investigated the probability of a successful projective measurement on the input state in the basis of the program register state by constructing a circuit that takes as inputs the states in the input and program registers. They showed that this probability increases with the number of copies of the state in the program register.  In particular, they proposed an implementation of a destructive state comparison test, where all output states are measured, with generic quantum states encoded in single-photons. They showed that the same implementation provides a programmable projective measurement scheme involving balanced beam splitters and photon number-resolving detectors, in which the states in the program registers approximate the direction of the measurement, which is performed on the state in the input register.

Their proposal, however, requires the creation and manipulation of high-dimensional superposition states, which is out of the reach of current experimental photonic technologies required for implementing quantum communication tasks. A major step in overcoming the difficulty in experimental realisation for such protocols requiring high-dimensional states was proposed by the theoretical work of Arrazola et al. \cite{arrazola2014coherent}: their work maps any protocol involving pure states of many qubits, unitary transformations and projective measurement to protocols based on coherent states of light in multiple optical modes, passive linear optical transformations and single-photon threshold detection. Since coherent states of light are natural realisations of states produced by lasers, these are highly efficient to produce and manipulate experimentally. This model was subsequently used to demonstrate quantum advantage in quantum communication tasks \cite{xu2015experimental,guan2016observation, kumar2019experimental}.\par

Motivated by the coherent state mapping of quantum protocols, we extend the results of~\cite{chabaud2018optimal} and introduce a programmable device which uses coherent states of light to perform a given projective measurement onto program coherent states, with commercially available passive linear optics components such as balanced beam splitters and single-photon threshold detectors. Our scheme takes as input a single-mode coherent state (\emph{the input} register) and $M-1$ copies of some coherent state $\ket{\beta}$ (the \emph{program} registers) and approximates the projective measurement  $\{\ket{\beta}\bra{\beta}, 1 - \ket{\beta}\bra{\beta}\}$ on the state in the input register in a single run. We provide and discuss three practical implementations of our programmable projective measurement scheme. The three schemes, which we will refer to as the \emph{Hadamard scheme}, the \emph{amplifier scheme}, and the \emph{looped amplifier scheme}, respectively, all provide an optimal projective measurement with one-sided error probability given a single copy of the input register and $M-1$ copies of the program registers, in the sense that they achieve the best possible approximation of the projective measurement using $M-1$ program states. The schemes differ, however, in the number of linear optics components.  While all three schemes are efficient, the \emph{looped amplifier scheme} is the most practical scheme requiring a single balanced beam splitter and a single threshold detector.

In addition to substantially reducing the experimental requirements, 
we obtain two additional advantages in our scheme compared to the original scheme of \cite{chabaud2018optimal}. First, our scheme is non-destructive, implying that the remaining output state after the projective measurement is not destroyed and can be used as a resource for subsequent tasks. The second advantage is that from the use of coherent states we obtain a more `faithful' projective measurement than using a single-photon encoding, implying better probability in carrying out a successful projective measurement.\par

Next, we extend our scheme to allow the input register to be obtained from an untrusted source: instead of requiring that the state in the input register is a coherent state, we allow any generic quantum state as an input, while the states in the program registers are still obtained from a trusted coherent state source. This setting is very natural in quantum cryptography and in verification of quantum state preparation \cite{amiri2017quantum, kumar2019practically}. In this setting, we also show an optimal approximate projective measurement on the input state, thus finding relevance when such a measurement is a part of some verification protocol. Our result enables an efficient verification of an unbounded untrusted source with only trusted coherent states, balanced beam splitters and threshold detectors.

As a final result, we give an application of our generalised scheme by showing an at-most quadratic improvement in an existing quantum fingerprinting protocol to approximate the Euclidean distance between two unit vectors  \cite{kumar2017efficient}. 

The paper is organised as follows. In section~\ref{singlecopy} we review the existing state comparison techniques for qubit states and coherent states. Following \cite{chabaud2018optimal}, we then consider the setting of a single input register state and multiple program register states, where both the input register state and the program register states are coherent states, in section~\ref{generaldis}. We introduce three different schemes for performing state comparison and programmable projective measurement with coherent states: the \emph{Hadamard scheme}, the \emph{amplifier scheme} and the \emph{looped amplifier scheme}. Further, in section~\ref{app:opti}, we give the proof for the optimality of our projective measurement for all three schemes, under the one-sided error requirement. We then analyse the robustness of our schemes by considering experimental imperfections in section~\ref{expimp}. In section~\ref{sec:genCV}, we drop the assumption that the incoming input register state is a coherent state.  We further prove that our projective measurement scheme is also optimal in this case. We conclude with section~\ref{fingerprint} by giving a concrete improvement of the quantum fingerprinting protocol to solve the Euclidean distance problem \cite{kumar2017efficient}.


\section{Quantum state comparison} \label{singlecopy}

\noindent A circuit for comparing two unknown qubit states, known as the swap test, was first introduced by Buhrman et. al.~\cite{buhrman2001quantum}. The analogue of this test when the states are unknown coherent states was introduced in \cite{arrazola2014coherent}. We briefly review these two tests here. 

\subsection{The swap test}

\begin{figure}
\includegraphics[scale=0.5]{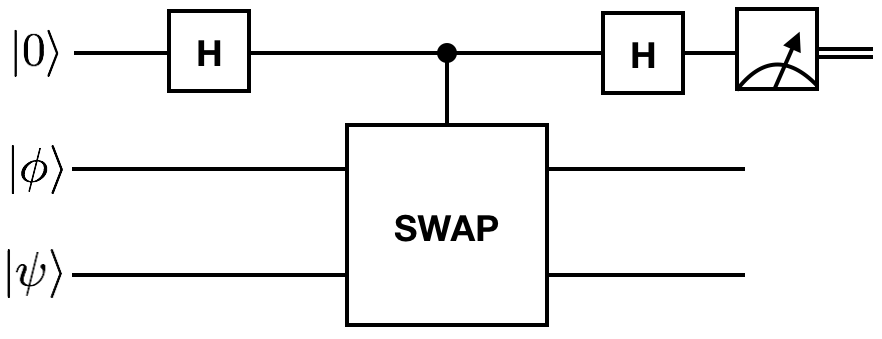}
\centering
\caption{Controlled-swap (C-SWAP) circuit employed by to compare the incoming qubit states. The ancilla qubit is measured in the computational basis and this relates to the probability of telling the two qubit states apart.}
\label{fig:Swaptest}
\end{figure}

\noindent The swap test uses a controlled-swap (C-SWAP) gate applied on two unknown qubit states $\ket{\phi}$ and $\ket{\psi}$, and controlled by an ancilla qubit, as shown in Fig.~\ref{fig:Swaptest}. Applying the circuit and measuring the ancilla qubit gives the output 1 with a probability $\mathbb{P}(1) =  \frac{1}{2}(1 - |\langle\phi|\psi\rangle|^2)$, and the output 0 with probability $\mathbb{P}(0)=1 - \mathbb{P}(1)$. \\

We define the completeness and soundness of this test as follows:\\

\noindent \textbf{Completeness}: If the states $\ket{\phi}$ and $\ket{\psi}$ are the same, then there is a zero probability of the outcome being $1$. We say that the test has perfect completeness $c_2=1$, where the completeness is defined as $c_2 = 1 - \mathbb{P}(1)$, when the input states are the same. The subscript denotes that two states have been used for testing. Alternatively, we say that the test meets the \textit{one-sided error requirement} when it has perfect completeness. \\

\noindent\textbf{Soundness}: If the states are different, then with finite probability $\mathbb{P}(1)$, one is able to tell the states apart. Thus the soundness, defined as $s_2 = \mathbb{P}(1)$, is strictly greater than $0$. The soundness of this scheme can be increased to any desired $1 - \delta$, by repeating the test $\mathcal O(\log\frac1\delta)$ times, using new copies of the states each time. \\
 
We note that this test provides an optimal comparison between two unknown states for the one-sided error probability. 

\subsection{Comparing two coherent states} \label{beamsplitter}

\begin{figure}
\includegraphics[scale=0.6]{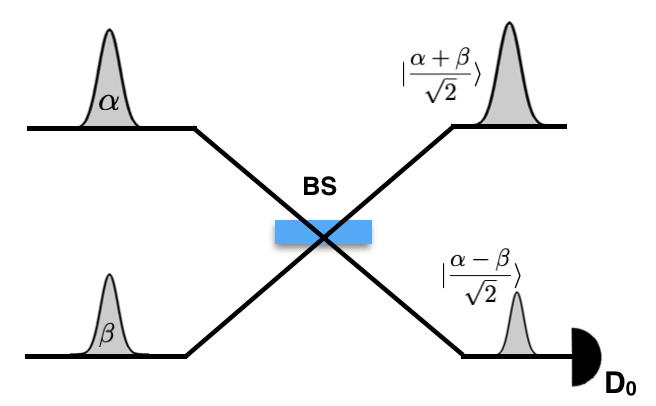}
\centering
\caption{Balanced beam splitter (BS) operation acting on input coherent states $\ket{\alpha}$ and $\ket{\beta}$. The lower output mode of the BS is measured with a single-photon threshold detector $D_0$. The probability of obtaining a click in $D_0$ relates to the projective measurement test of distinguishing the two coherent states.}
\label{fig:OpticalSwaptest}
\end{figure}

\noindent The above swap test compares two unknown qubit states. If the unknown states are coherent states instead, then an analogous test can be performed by simply mixing the states on a balanced beam splitter (BS) and observing a photon click with a single-photon threshold detector (detector $D_0$ in Fig.~\ref{fig:OpticalSwaptest}).

This can be seen as follows. The beam splitter transforms the input mode creation operators $\{\hat{a}^{\dagger},\hat{b}^{\dagger}\}$ into the output mode creation operators $\{\hat{c}^{\dagger},\hat{d}^{\dagger}\}$. This input to output conversion is given by:
\begin{equation}
\begin{split}
\hat{a}^{\dagger} & \rightarrow \frac{1}{\sqrt{2}}(\hat{c}^{\dagger}+\hat{d}^{\dagger}), \\
\hat{b}^{\dagger} & \rightarrow \frac{1}{\sqrt{2}}(\hat{c}^{\dagger}-\hat{d}^{\dagger}).
\end{split}
\end{equation} 
The input state at the beam splitter is
\begin{equation}
\ket{\alpha}_a\otimes \ket{\beta}_b,
\end{equation}
where the subscripts denote the mode in which the coherent states enter the beam splitter.
In the absence of experimental imperfections, this yields the output state
\begin{equation}\label{output}
\left|\frac{\alpha + \beta}{\sqrt{2}}\right\rangle_c\otimes \left|\frac{\alpha - \beta}{\sqrt{2}}\right\rangle_d.
\end{equation}
The probability of obtaining a click in the detector $D_0$ (mode $d$) is
\begin{equation}
\mathbb{P}_{D_0} = 1 - \exp\left(-\frac{|\alpha - \beta|^2}{2}\right) = 1 - |\langle\alpha|\beta\rangle|.
\label{D1}
\end{equation}
We can relate the completeness $c_2$ and soundness $s_2$ of the test to the trace distance of the tested states $\ket{\alpha}$ and $\ket{\beta}$. The trace distance for two coherent states $\{\ket{\alpha}, \ket{\beta}\}$ is
\begin{equation}
\|\ket{\alpha}\bra{\alpha} - \ket{\beta}\bra{\beta}\|_{\text{tr}} = \sqrt{1 - |\langle\alpha|\beta\rangle|^2} = \sqrt{1 - e^{-|\alpha - \beta|^2}}.
\label{eq:tracedistance2}
\end{equation}
We assign the detection event (obtaining a click in $D_0$) the value 1, and to the other detection event (no click in detector $D_0$) the value 0. The completeness and soundness for this test are:\\

\noindent \textbf{Completeness}: If the states are the same, then their trace distance is $0$, since $\ket\alpha = \ket\beta$. This implies that $|\braket{\alpha|\beta}| = 0$ thus leading to $\mathbb{P}_{D_0} = 0$.  This ensures perfect completeness $c_2 = 1$, where the subscript denotes the size of the interferometer. \\

\noindent \textbf{Soundness}:  Suppose the states $\ket\alpha$ and $\ket\beta$ are $\epsilon$ far in trace distance meaning,
\begin{equation}
\|\ket{\alpha}\bra{\alpha} - \ket{\beta}\bra{\beta}\|_{\text{tr}} \geq \epsilon \implies   |\braket{\alpha|\beta}| \leq \sqrt{1 - \epsilon^2}
\end{equation}
 From this we can upper bound the soundness $s_2$ as, 
 \begin{equation}
   s_2 =  \mathbb{P}_{D_0} \geq 1 - \sqrt{1 - \epsilon^2} 
 \end{equation}
This implies that the soundness is strictly greater than 0 for a non zero $\epsilon$.  The soundness can be decreased to any desired $\delta$ by repeating the measurement procedure for $\mathcal O(\log\frac1\delta)$ runs. Further, this method provides an optimal comparison between two unknown coherent states under one-sided error probability. We prove this more generally in section~\ref{app:opti}
%


\section{Generalised single run coherent state comparison} \label{generaldis}

\noindent Having briefly reviewed existing unknown state comparison techniques, we now consider the scenario where we have a single copy of an unknown coherent state $\ket{\alpha}$ in the input register and multiple copies of the coherent state $\ket{\beta}$ in the program register. The task is to check if the state in the input register is equal to the state in the program register. 

In the simplest case, the state comparison can be performed with a single copy of the state in the program register, like in the previous section. This succeeds with a probability given by Eq.~(\ref{D1}). In this section, we prove that having multiple copies of $\ket{\beta}$ increases the success probability of state comparison with the state $\ket{\alpha}$. For this, we first provide a generalised interferometer construction, the \emph{Hadamard scheme}, based on Hadamard-Walsh transforms, following~\cite{chabaud2018optimal}. We then derive the \emph{amplifier scheme} requiring much less optical gates and detectors than the previous scheme. Finally, we modify the \emph{amplifier scheme} to construct an even simpler scheme, the \emph{looped amplifier scheme}.

\subsection{The Hadamard scheme}

\noindent In~\cite{chabaud2018optimal}, it is shown how to perform state comparison and programmable measurements with linear optics using generic quantum states encoded in degrees of freedom of single photons. These photons are fed into an specific interferometer, the Hadamard interferometer, which we review below. All output modes of the interferometer are then measured with photon number-resolving detectors and the classical outcomes are post-processed to retrieve the statistics of a projective measurement. In what follows, we adapt and simplify their approach to the case where the input states are coherent states. We show that single-photon threshold detectors are sufficient for our needs, obtaining a practical scheme. \\

\begin{figure}
\begin{center}
\includegraphics[scale=0.4]{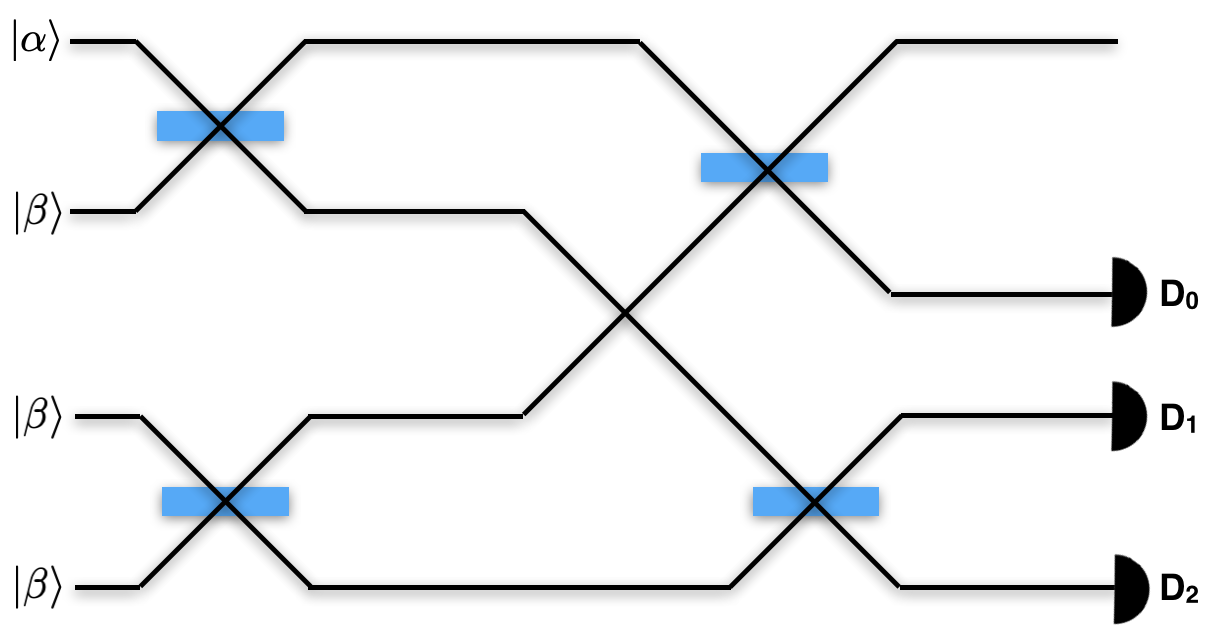}
\caption{Hadamard interferometer with 4 input modes.The input states are one input register state $\ket{\alpha}$ and three local states $\ket{\beta}$, one in each mode. The detectors $D_i$ are single-photon threshold detectors, $\forall i \in \{0,2\}$.}
\label{fig:4swap}
\end{center}
\end{figure}

\noindent \textbf{Input State}: Suppose the input is $M$ coherent states, where $M$ is a power of $2$: 
\begin{equation}
\ket{\alpha}_0\otimes\ket{\beta}_1\dots\otimes\ket{\beta}_{M-1},
\end{equation}
where the subscript denotes the mode in which the coherent state enters the generalised interferometer (indexed from $0$ to $M-1$). For brevity, we address this state as $\ket{\alpha\beta\dots\beta}$.
This input state is then fed in an interferometer of size $M$. For $M=4$ spatial modes, this interferometer is described by the Hadamard-Walsh transform of order $2$:
\begin{equation}
H_2 := H^{\otimes2} = \frac{1}{\sqrt{2}}\begin{pmatrix} H & H \\ H & -H \end{pmatrix},
\end{equation}
where $H$ is a Hadamard matrix. The corresponding interferometer for $M = 4$ is depicted in Fig.~\ref{fig:4swap}.

In the general case, the Hadamard interferometer of order $M$ is described by the Hadamard-Walsh transform of order $n=\log M$, which is defined by:
\begin{equation}
H_{n}:= H^{\otimes n},
\label{interferometern}
\end{equation}
with $H_0=1$ and $H_1=H$. 
\\\\
\textbf{Output state}: The input coherent states $\ket{\alpha\beta\dots\beta}$ upon interaction with the interferometer of order $n$ transforms as:
\be
\begin{aligned}
\ket{\alpha\beta\dots\beta}\mapsto H_n\ket{\alpha\beta\dots\beta}&=\ket{\delta_0\delta_1\dots\delta_{M-1}},
\end{aligned}
\ee
where, with a simple induction, we obtain $\delta_0=\frac{\alpha+(M-1)\beta}{\sqrt{M}}$ and $\delta_k=\frac{\alpha-\beta}{\sqrt{M}}$ for $k>0$. Thus the last $M-1$ modes have the same probability of a click when we detect with single-photon threshold detectors. The probability $\mathbb{P}_\emptyset$ that none of the $M-1$ detectors clicks is:
\be
\begin{aligned}
\mathbb{P}_\emptyset(\alpha,\beta,M)&=\prod_{k=1}^{M-1}{[1-\mathbb{P}(\text{click in $k^{th}$ mode})]}\\
&=\prod_{k=1}^{M-1}{[1-(1 - \exp(-|\delta_k|^2))]}\\
&=\exp\left(-\frac{M-1}{M}|\alpha-\beta|^2\right)\\
&=(|\!\braket{\alpha|\beta}\!|^2)^{1-\frac{1}{M}}.
\end{aligned}
\label{Eq:Noclick}
\ee
In particular, for all $\alpha,\beta\in\mathbb{C}$, $\mathbb{P}_{\emptyset}(\alpha,\beta,+\infty)=|\!\braket{\alpha|\beta}\!|^2$, which corresponds to a perfect projective measurement of the states $\ket{\alpha}$ and $\ket{\beta}$.  Writing $x:=|\!\braket{\alpha|\beta}\!|^2$ the overlap of the test and program register states, we obtain
\be
\mathbb{P}_\emptyset(x,M)=x^{1-\frac{1}{M}}.
\ee
Assigning to this detection event (none of the detectors clicks) the value 0, and to other detection events (at least one of the $M-1$ detectors clicks) the value 1, we obtain a device whose statistics mimic those of a projective measurement, with
\be
\mathbb{P}_M(0)=1-\mathbb{P}_M(1)=x^{1-\frac{1}{M}}.
\label{probaCS}
\ee
The test based on single-photon encoding from~\cite{chabaud2018optimal} requires using $M$ number-resolving detectors. On contrary, the encoding with coherent states requires $M-1$ single-photon threshold detectors. Experimentally, this is much more cost effective and relatively easier to implement. The test based on coherent state encoding has the following characteristics:\\

\noindent \textbf{Completeness}: If the states are the same, then the trace distance $\|\ket{\alpha}\bra{\alpha} - \ket{\beta}\bra{\beta}\|_{\text{tr}} = 0$, and hence the probability of having the detection event 1 is 0. Thus, the completeness of this scheme is $c_M = 1$.
\\

\noindent \textbf{Soundness}: If the states $\{\ket{\alpha}, \ket{\beta}\}$ are $\epsilon$ far apart in trace distance, then
using Eq.~(\ref{eq:tracedistance2}), and the fact that the soundness is $s_M = 1 - \mathbb{P}_\emptyset(\alpha,\beta,M)$ we obtain,
\be
 s_M \geq 1 - (1 - \epsilon^2)^{1 - \frac{1}{M}}
\ee
\noindent Moreover, contrary to single-photon encoding, the Hadamard scheme with coherent state encoding is non-destructive, as the first output mode is not measured.

An other advantage with coherent state encoding is that it gives a more faithful projective measurement than the single-photon encoding in~\cite{chabaud2018optimal}. Indeed,the statistics corresponding to a perfect projective measurement are
\be
\mathbb{P}(0)=1-\mathbb{P}(1)=x,
\label{probaperfect}
\ee
where $x$ is the overlap between the input state and the program state, while for the single-photon encoding, with an $M$-mode input state $\ket{\phi\psi\dots\psi}$, the corresponding statistics as obtained in~\cite{chabaud2018optimal} are
\be
\mathbb{P}_M(0)=1-\mathbb{P}_M(1)=\frac{1}{M}+\left(1-\frac{1}{M}\right)x,
\label{probaSP}
\ee
and for any given value of the overlap $x$ we have:
\be
x\le x^{1-\frac{1}{M}}\le\frac{1}{M}+\left(1-\frac{1}{M}\right)x,
\ee
where the term on the left hand side corresponds to a perfect projective measurement (Eq.~\ref{probaperfect})), the central term corresponds to our coherent state scheme  (Eq.~(\ref{probaCS})) and the term on the right hand side corresponds to the single-photon scheme (Eq.~(\ref{probaSP})).
\begin{figure}
\begin{center}
\includegraphics[scale=0.3]{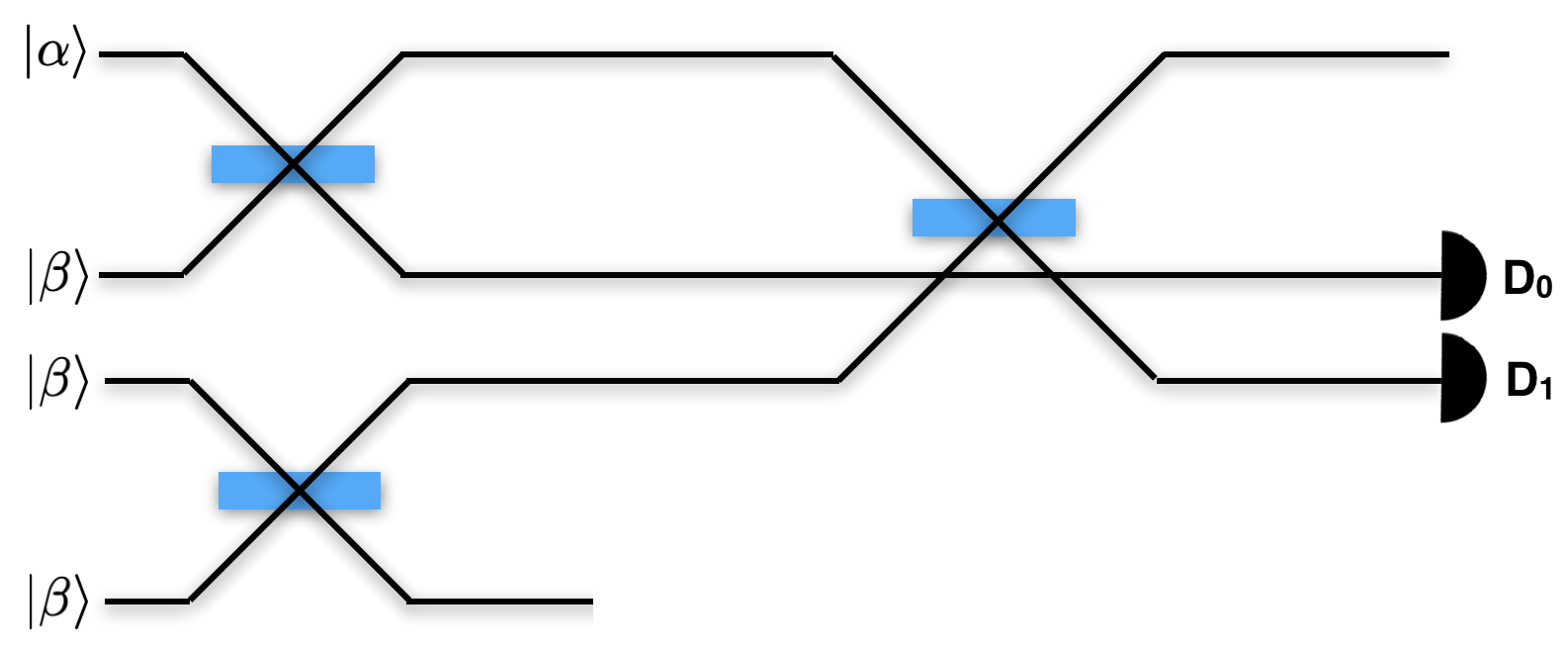}
\caption{Amplifier scheme with 4 input modes. The input states are one tested state $\ket{\alpha}$ and three local states $\ket{\beta}$, one in each mode. The detectors $D_i$ are single-photon threshold detectors. This scheme may be contrasted with the one in Fig.~\ref{fig:4swap}}.
\label{fig:simpl}
\end{center}
\end{figure}
In particular, for a given size $M$, the maximal statistical gap with a perfect projective measurement is,
\be
\begin{aligned}
e_{SP}(M)&=\max_{x\in[0,1]}{\left|\left[\frac{1}{M}+\left(1-\frac{1}{M}\right)x\right]-x\right|}\\
&=\frac{1}{M},
\end{aligned}
\ee
for the single-photon encoding, and
\be
\begin{aligned}
e_{CS}(M)&=\max_{x\in[0,1]}{\left|\left( x^{1-\frac{1}{M}}\right)-x\right|}\\
&=\frac{(M-1)^{M-1}}{M^M}\\
&\sim\frac{1}{e}\cdot\frac{1}{M},
\end{aligned}
\ee
for the coherent state encoding, which is lower than the single-photon encoding gap.
This happens because for the single-photon encoding no assumption is made about the states $\ket\phi$ and $\ket\psi$, while the states $\ket\alpha$ and $\ket\beta$ are assumed to be coherent states. This additional information about the states allows to better approximate a perfect projective measurement with the same number of input states. A related question would be, is the generalised Hadamard scheme optimal? Or can a better measurement setting improve the state comparison? We show in section~\ref{app:opti} that the generalised Hadamard interferometer is actually optimal for approaching perfect projective measurement with coherent states, under the one-sided error requirement. However, we already show in the next paragraph that there exists a simpler measurement setting than the Hadamard interferometer, achieving the same performance in the test.

\begin{figure*}[t]
    \begin{center}
        \includegraphics[scale=0.5]{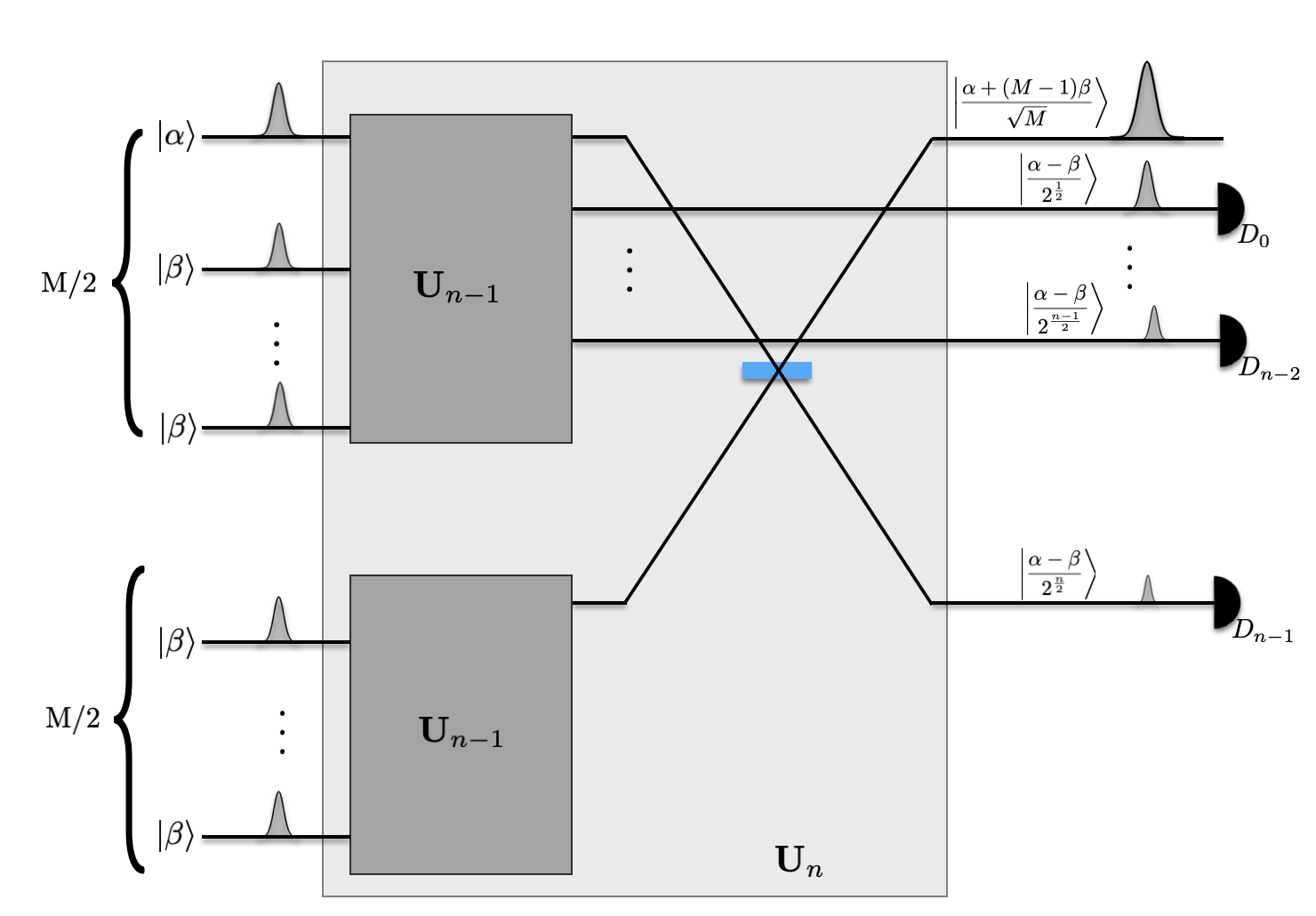}
        \caption{General amplifier scheme of size $M$, with one copy of $\ket{\alpha}$ and $M - 1$ copies of $\ket{\beta}$: the first output modes of two interferometers described by $U_{n-1}$ are mixed on a balanced beam splitter. The input states are one tested state $\ket{\alpha}$ and $M-1$ local states $\ket{\beta}$, one in each mode. Indexing the spatial modes from $0$ to $M-1$, the $2^k$ output modes are measured with single-photon threshold detectors labelled $D_k$, for $k = 0\dots n-1$. }
        \label{fig:induction}
    \end{center}
\end{figure*}
\subsection{The amplifier scheme} \label{ampscheme}

\noindent The Hadamard scheme of size $M$ described in the previous section uses $(M\log M)/2$ balanced beam splitters and $M-1$ single-photon threshold detectors. We introduce a simplified scheme of the same size, which only uses $M-1$ balanced beam splitters and $\log M$ detectors, and show that it achieves the same performance than the Hadamard scheme. We refer to this scheme as the \textit{amplifier} scheme, since it maps identical input coherent states to an amplified coherent state in the first output mode and the vacuum in all other modes.

For $M=4$ spatial modes, this interferometer acting on modes $\{0,1,2,3\}$ is described by the following unitary matrix:
\be
\ba
U_2&=\begin{pmatrix}\frac{1}{2}&\frac{1}{2}&\frac{1}{2}&\frac{1}{2}\\ \frac{1}{\sqrt2}&-\frac{1}{\sqrt2}&0&0\\ \frac{1}{2}&\frac{1}{2}&-\frac{1}{2}&-\frac{1}{2}\\0&0&\frac{1}{\sqrt2}&-\frac{1}{\sqrt2}\end{pmatrix}\\
&=\begin{pmatrix}\frac{1}{\sqrt2}&0&\frac{1}{\sqrt2}&0\\0&1&0&0\\ \frac{1}{\sqrt2}&0&-\frac{1}{\sqrt2}&0\\0&0&0&1\end{pmatrix}\times\begin{pmatrix}\frac{1}{\sqrt2}&\frac{1}{\sqrt2}&0&0\\ \frac{1}{\sqrt2}&-\frac{1}{\sqrt2}&0&0\\0&0&\frac{1}{\sqrt2}&\frac{1}{\sqrt2}\\0&0&\frac{1}{\sqrt2}&-\frac{1}{\sqrt2}\end{pmatrix}\\
&=H_{0,2}\times(H_{0,1}\oplus H_{2,3}),
\ea
\label{U2}
\ee
where $H_{i,j}$ corresponds to the balanced beam splitter operation acting on modes $i$ and $j$ (where the modes are indexed from $0$ to $M-1$) and identity on the other modes (Fig.~\ref{fig:simpl}). 

The generalised amplifier interferometer is defined by induction:
\be
U_n=H_{0,M/2}\times(U_{n-1}\oplus U_{n-1}),
\label{U}
\ee
where $n=\log M$ and where $U_1=H_{0,1}=H$ is a Hadamard matrix. This induction relation is illustrated in Fig.~\ref{fig:induction}.
Indexing the spatial modes from $0$ to $M-1$, the $2^k$ output modes are measured with single-photon threshold detectors, for $k=0\dots n-1$. A simple induction shows that the output state in the $2^k$ output mode is $|\frac{\alpha-\beta}{2^{\frac{k+1}{2}}}\rangle$.
Hence, the probability that none of the $n=\log M$ detectors clicks is given by
\be
\begin{aligned}
P_\emptyset(\alpha,\beta,M)&=\prod_{k=0}^{n-1}{[1-\mathbb{P}(\text{click in the }2^kth\text{ mode})]}\\
&=\prod_{k=0}^{n-1}{\left[1-\left(1 - \exp\left(-\left|\frac{\alpha-\beta}{2^{\frac{k+1}{2}}}\right|^2\right)\right)\right]}\\
&=\exp\left(-\sum_{k=0}^{n-1}{\left(\frac{1}{2}\right)^{k+1}}|\alpha-\beta|^2\right)\\
&=\exp\left(-\frac{M-1}{M}|\alpha-\beta|^2\right)\\
&=(|\!\braket{\alpha|\beta}\!|^2)^{1-\frac{1}{M}},
\end{aligned}
\label{Eq:Noclick2}
\ee
thus retrieving the statistics obtained with the Hadamard scheme, using only $n=\log M$ detectors. Moreover, another simple induction shows that the amplifier interferometer can be implemented with only $M-1$ balanced beam splitters.

\subsection{Looped amplifier scheme}

\begin{figure*}[t]
\begin{center}
\includegraphics[scale=0.6]{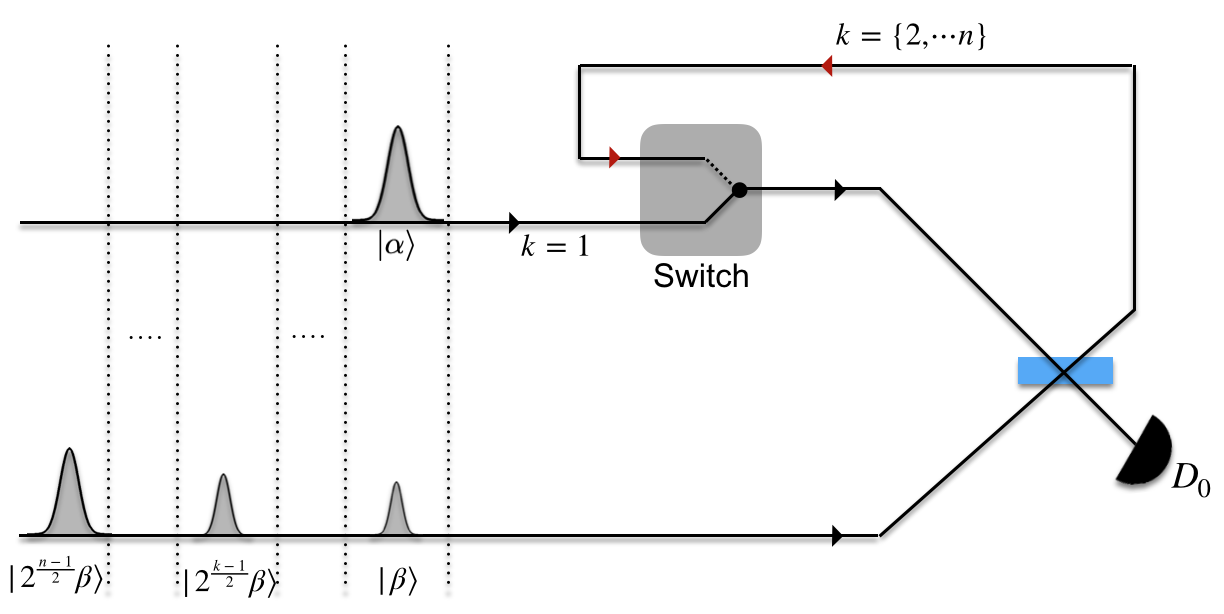}
\caption{The looped amplifier scheme. The coherent state $\ket\alpha$ is sent once, while at each loop a new coherent state $\ket{2^{(k-1)/2}}\beta$ for $k = \{1,.. n\}$, with $n = \log M$. The switch ensures a closed loop with the arm corresponding the the red arrow after the input state $\ket\alpha$ passes through. The clicks are collected in the detector $D_0$ for all the $n$ instances of the beam-splitter interaction. In the ideal case, when $\ket\alpha = \ket\beta$, the detector $D_0$ does not click across any of the $n$ instances. If the two states are however different, there is a finite probability of a click across the $n$ instances.}
\label{fig:loop}
\end{center}
\end{figure*}

\noindent Noting the recursive character of the amplifier scheme, we present another possible implementation of the amplifier scheme using a looped beam splitter interaction, one single-photon threshold detector and an active optical element, namely a switch (Fig.~\ref{fig:loop}).

This setup now uses an active optical element and a constant number of passive linear optical elements, and approximates a perfect projective measurement up to arbitrary precision. It works in the following manner. Initially the states $\ket\alpha$ and $\ket\beta$ are mixed on a balanced beam splitter. This results in a probability of not obtaining a click in the detector $D_0$ given by
\be
P_\emptyset(\alpha,\beta,1) = |\braket{\alpha|\beta}|
\ee
If a click is detected in $D_0$, then one immediately concludes that $\ket\alpha \neq \ket\beta$. Otherwise, in the next iteration, the switch connects to the upper arm and the next interaction with the beam splitter results in,
\be
\Ket{\frac{\alpha + \beta}{\sqrt{2}}} \otimes \ket{\sqrt{2}\beta} \rightarrow \Ket{\frac{\alpha + 3\beta}{2}}\otimes \Ket{\frac{\alpha - \beta}{2}}_{D_0},
\ee
where in the lower arm the amplitude of the new coherent pulse has been multiplied by $\sqrt2$. The coherent state $\ket{\sqrt{2}\beta}$ is produced using variable optical attenuator (VOA): an active optical element to prepare coherent states with desired amplitude by the varying the intensity in the attenuator. 
Iterating this over $n = \log M$ runs, where at the $k$-th run, the upper arm corresponding the unmeasured output state interacts with the local state $\ket{2^{(k-1)/2}\beta}$ (produced using VOA) in the beam splitter, the probability that there is no click obtained $D_0$ over all the runs is the same as Eq.~(\ref{Eq:Noclick2}). Hence, by construction, the statistics of the setup after $n-1$ pulses $\{\ket\beta,\cdots \ket{2^{(n-1)/2}\beta}\}$ sent reproduce those of the amplifier scheme of size $M$.

The three schemes discussed provide experimentally friendly devices to perform a variety of quantum information processing tasks using coherent states, ranging from state comparison to programmable projective measurements, in a non-destructive manner. We show in the next section that these schemes are optimal for coherent state comparison under the one-sided error requirement.


\section{Optimality of the generalised test}
\label{app:opti}

\noindent An extension of the results in~\cite{sedlak2007unambiguous}, which studies the problem of unambiguous comparison of unknown coherent states, proves the optimality of both the Hadamard scheme and the amplifier scheme for coherent states comparison, under the one-sided error requirement. In other terms, since these schemes satisfy the promise of perfect completeness, we show that they achieve minimal soundness.

\subsection{Optimal POVM for comparing coherent states under the one-sided error requirement}

\noindent Let $\{\Pi_0,\Pi_1\}$ be a POVM for comparing coherent states $\ket\alpha$ and $\ket\beta$ under the one-sided error requirement, when provided a single copy of $\ket\alpha$ and $M-1$ copies of $\ket\beta$ (the proof of~\cite{sedlak2007unambiguous} assumes $M=2$). The operator $\Pi_0$ corresponds to saying that the states $\ket\alpha$ and $\ket\beta$ are the same, while the operator $\Pi_1$ corresponds to saying that they are different. These operators thus verify the following conditions:
\begin{equation}
\Pi_0,\Pi_1\ge0,\text{ }\Pi_0+\Pi_1=\mathbb{I},
\label{cond0}
\end{equation}
where $\mathbb{I}$ is the identity operator, and
\be
\forall\alpha\in\mathbb{C},\text{ Tr}\left[\Pi_1\ket\alpha\bra\alpha^{\otimes M}\right]=0,
\label{cond1}
\ee
where the last condition is the one-sided error requirement. Integrating this condition over $\mathbb{C}$ yields
\begin{equation}
0=\int{d^2\alpha\text{Tr}\left[\Pi_1\ket\alpha\bra\alpha^{\otimes M}\right]}=\text{Tr}\left[\Pi_1\Delta_M\right],
\label{cond2}
\end{equation}
where we have defined
\begin{equation}
\Delta_M=\int{d^2\alpha\ket\alpha\bra\alpha^{\otimes M}}\ge0.
\label{Delta1}
\end{equation}
Note that the condition in~(\ref{cond2}) is equivalent to the one-sided requirement in~(\ref{cond1}) because the operators $\Pi_1$ and $\ket\alpha\bra\alpha^{\otimes M}$ are positive. 

The operator $\frac{M}{\pi}\Delta_M$ is actually a projector. This result can be obtained by writing the state $\ket\alpha$ in the Fock basis and an integration in polar coordinates, where $\alpha = re^{i\theta}$ (the derivation is detailed in Appendix~\ref{app:Delta}). From Eq.~(\ref{Delta1}) we obtain
\begin{equation}
\Delta_M=\frac{\pi}{M}\sum_{N=0}^{\infty}{\ket{\chi_N^M}\bra{\chi_N^M}},
\label{Delta2}
\end{equation}
where we have defined for all $N\ge0$,
\begin{equation}
\ket{\chi_N^M}:=M^{-N/2}\sum_{\sum_j k_j=N}{\sqrt{\frac{N!}{k_1!\dots k_M!}}\ket{k_1\dots k_M}}.
\label{chi}
\end{equation}
With the multinomial formula, we obtain $\braket{\chi_N^M|\chi_N^M}=1$ for all $N\ge0$, and since the states $\ket{\chi_N^M}$ have exactly $N$ photons, we have $\braket{\chi_N^M|\chi_{N'}^M}=\delta_{N,N'}$ for all $N,N'\ge0$. The states $\ket{\chi_N^M}$ thus are orthonormal and with Eq.~(\ref{Delta2}), the operator $\frac{M}{\pi}\Delta_M$ is a projector.

By Eq.~(\ref{cond2}), the supports of $\Pi_1$ and $\frac{M}{\pi}\Delta_M$ are disjoint, and by Eq.~(\ref{cond0}) we see that $\Pi_0+\Pi_1=\mathbb{I}$, so the support of $\frac{M}{\pi}\Delta_M$ is included in the support of $\Pi_0$. The optimal POVM $\{\Pi_0^{opt},\Pi_1^{opt}\}$ for state comparison minimises the error probability, hence with the one-sided error requirement $\Pi_0^{opt}$ must have minimal support, meaning that
\begin{equation}
\Pi_0^{opt}=\frac{M}{\pi}\Delta_M=\sum_{N=0}^{+\infty}{\ket{\chi_N^M}\bra{\chi_N^M}}\quad\text{and}\quad\Pi_1^{opt}=\mathbb{I}-\Pi_0^{opt}.
\label{POVMopt}
\end{equation}
Note that, with the same proof, this choice of POVM is also optimal in the generalised setting where one is given one unknown generic state (not necessarily a coherent state) and $M-1$ unknown coherent states, and is asked to test if all the states are identical or not.\\

\subsection{The Hadamard interferometer is optimal for coherent state comparison}

\noindent We now show that the POVM $\{\Pi_0^h,\Pi_1^h\}$ corresponding to the Hadamard interferometer with threshold detection of the last $M-1$ modes is optimal for coherent state comparison under the one-sided error requirement, i.e.,
\be
\Pi_0^h=\Pi_0^{opt},
\ee
where $\Pi_0^{opt}$ is defined in Eq.~(\ref{POVMopt}). We have
\be
\Pi_0^h=\hat H_n^\dag\Pi_0^d\hat H_n,
\ee
where $\hat H_n$ is the unitary evolution corresponding to the action of the interferometer of order $M$ defined in Eq.~(\ref{interferometern}) in the $M$-mode infinite-dimensional Hilbert space, with $n=\log M$, and where
\be
\Pi_0^d=\mathbb{I}\otimes\ket0\bra0^{\otimes(M-1)}
\ee
is the POVM operator corresponding to the event where none of the $M-1$ threshold detectors clicks. We obtain
\begin{equation}
\begin{aligned}
\Pi_0^h&=\hat H_n^\dag\left(\mathbb{I}\otimes\ket0\bra0^{\otimes(M-1)}\right)\hat H_n\\
&=\sum_{N=0}^{+\infty}{\hat H_n^\dag\left(\ket{N}\bra{N}\otimes\ket0\bra0^{\otimes(M-1)}\right)\hat H_n}.
\end{aligned}
\label{Pi0h}
\end{equation}
For $k=1,\dots,M$, we write $\hat a_k^\dag$ the creation operator for the $k^{th}$ mode. For all $N\ge0$ we have,
\begin{widetext}
\begin{equation}
\begin{aligned}
\hat H_n^\dag\left(\ket{N}\otimes\ket0^{\otimes(M-1)}\right)&=\frac{1}{\sqrt{N!}}\hat H_n^\dag (\hat a_1^\dag)^N\ket0^{\otimes M}\\
&=\frac{1}{\sqrt{N!}}(\hat H_n^\dag \hat a_1^\dag \hat H_n)^N\ket0^{\otimes M}\\
&=\frac{M^{-N/2}}{\sqrt{N!}}(\hat a_1^\dag+\dots+\hat a_M^\dag)^N\ket0^{\otimes M}\\
&=\frac{M^{-N/2}}{\sqrt{N!}}\sum_{k_1+\dots+k_M=N}{\frac{N!}{k_1!\dots k_M!}(\hat a_1^\dag)^{k_1}\dots(\hat a_M^\dag)^{k_M}}\ket0^{\otimes M}\\
&=M^{-N/2}\sum_{k_1+\dots+k_M=N}{\sqrt{\frac{N!}{k_1!\dots k_M!}}\ket{k_1\dots k_M}}\\
&=\ket{\chi_N^M},
\end{aligned}
\end{equation}
\end{widetext}
where we have used $\hat H_n\ket0^{\otimes M}=\ket0^{\otimes M}$, $\hat H_n^\dag \hat H_n=\mathbb I$, $\hat H_n^\dag \hat a_1^\dag \hat H_n=\frac{\hat a_1^\dag+\dots+\hat a_M^\dag}{\sqrt M}$, the multinomial formula, and Eq.~(\ref{chi}). With Eqs.~(\ref{POVMopt}) and (\ref{Pi0h}), this concludes the proof.

Given that the statistics obtained with the amplifier scheme and the looped amplifier scheme mimic those of the Hadamard scheme, these schemes are also optimal for the same state comparison task.


\section{Analysis with Experimental Imperfections for $M=4$ modes} \label{expimp}

\noindent While these devices are relatively easy to implement, any implementation would suffer from experimental imperfections. In this section, we analyze the performance of the amplifier scheme in presence of such imperfections. Our error model is the following, with three major sources of error: $(i)$ limited detector efficiency and channel transmission loss, characterized by a parameter $0\leq\eta\leq1$. This changes the coherent state $\alpha$ to $\sqrt{\eta}\alpha$ thus reducing the probability of obtaining a click using a single-photon threshold detector by a factor $\eta$; $(ii)$ limited beam splitter visibility $0\leq\nu\leq1$, which may lead to a click in the wrong detector; $(iii)$ dark counts in the detectors characterized by a probability $p_{dark}$. For our analysis, the click probability due to the coherent states is $\mathcal O(1)$ and thus significantly larger than the dark count probability $p_{dark}$ $(\sim 10^{-8})$ observed in the standard commercially available superconducting nano-wire single photon detectors. The dark count effect can thus be safely ignored.

For $M=2$, when the input $\ket{\alpha}$, $\ket{\beta}$ is fed in an imperfect beam splitter, the transformation from input modes to output modes is the following:
\begin{equation}
\small
\ket{\alpha}\otimes\ket{\beta}\mapsto\ket{\sqrt\nu k_{+} +\sqrt{1-\nu}k_{-}}\otimes\ket{\sqrt\nu k_{+}+\sqrt{1-\nu} k_{-}}.
\end{equation}
where $k_{+} = (\alpha + \beta)/\sqrt{2}$ and $k_{-} = (\alpha - \beta)/\sqrt{2}$.
The corresponding unitary transformation is
\begin{equation}
H'=\frac{1}{\sqrt2}\begin{pmatrix} A&B\\A&-B \end{pmatrix},
\label{Eq:hadamard}
\end{equation}
where $A = \sqrt{\nu} + \sqrt{1 - \nu}$, and $B = \sqrt{\nu} - \sqrt{1 - \nu}$.


Next, we consider the case of $M = 4$ spatial modes (Fig~\ref{fig:simpl}), indexed from $0$ to $3$. We apply the imperfect transformation on the input $\ket{\alpha\beta\beta\beta}$. This results in
\be
\begin{aligned}
\ket{\alpha\beta\beta\beta}\mapsto U'_2\ket{\alpha\beta\beta\beta}&=\ket{\delta_0\delta_1\delta_2\delta_3},
\end{aligned}
\ee
where from Eq.~(\ref{U2}) we derive
\be
\ba
U'_2&=H'_{0,2}\times(H'_{0,1}\oplus H'_{2,3})\\
&=\begin{pmatrix}\frac{1}{2}A^2&\frac{1}{2}AB&\frac{1}{2}AB&\frac{1}{2}B^2\\ \frac{1}{\sqrt2}A&-\frac{1}{\sqrt2}B&0&0\\ \frac{1}{2}A^2&\frac{1}{2}AB&-\frac{1}{2}AB&-\frac{1}{2}B^2\\0&0&\frac{1}{\sqrt2}A&-\frac{1}{\sqrt2}B\end{pmatrix},
\ea
\ee
with $A=\sqrt\nu+\sqrt{1-\nu}$ and $B=\sqrt\nu-\sqrt{1-\nu}$.
We obtain $\delta_1=\frac{A\alpha-B\beta}{\sqrt2}$, and $\delta_2 = \frac{A^2\alpha-B^2\beta}{2}$. 
Adding the channel and detector losses $\eta$, the output is mapped as $\delta_k \mapsto\sqrt{\eta}\delta_k$, for all $k$.

Similar to the analysis without experimental imperfections, the output modes $1$ and $2$ of the imperfect amplifier interferometer are measured, with the coherent state input being $\ket{\alpha\beta\beta\beta}$. The probability $\mathbb{P}_{\emptyset}$ that none of the two detectors clicks is:
\be
\mathbb{P}_\emptyset(\alpha,\beta,\nu,\eta,M=4)=\exp\left(-\eta(|\delta_1|^2 + |\delta_2|^2)\right).\\
\ee
Assigning to the detection event \textit{no detector clicks} the value 0, and to other detection events, i.e., \textit{at least one of the detectors clicks}, the value 1, we obtain a device whose statistics mimic those of a projective measurement.
\\\\
\textbf{Completeness}: When the states are the same, the correctness, which is the probability of not obtaining  the detection event 1 is
\begin{equation}
c_4^{exp} = \mathbb{P}_\emptyset(\alpha,\alpha,\nu,\eta,4) =  \exp(-2\eta(1-\nu)(1+2\nu)|\alpha|^2).
\label{4completeexp}
\end{equation}
We observe that if $\nu = 1$ (no imperfections), then $c_4^{exp} = 1$, thus we obtain perfect completeness. 
\\\\
\textbf{Comparison of Completeness with the $M=2$ scheme:}
The analogous completeness in $M = 2$ scheme is
\begin{equation}
c_2^{exp} = \mathbb{P}_\emptyset(\alpha,\alpha,\nu,\eta,2) =  \exp(-2\eta(1 - \nu)|\alpha|^2).
\label{2completeexp}
\end{equation}
From Eq.~(\ref{2completeexp}) and Eq.~(\ref{4completeexp}), we observe that $c_2^{exp} \le c_4^{exp}$, which implies that the completeness in the $M = 4$ scheme is less than the completeness in $M = 2$ scheme. The reduction in completeness probability for the $M = 4$ scheme is precisely what accounts for an increase in soundness probability (when the input and program register states are different), which we detail in the next paragraph.
\\\\
\textbf{Soundness}: If the states are different, the probability of obtaining the detection event 1 (soundness) is computed in Appendix~\ref{app:s4} and given by:

\begin{widetext}
\begin{equation}
\small
s_4^{exp}=1 - \exp\left[(4\nu^2-1)|\alpha-\beta|^2+4\left[(1+2\nu)(1-\nu)+2\sqrt{\nu(1-\nu)}\right]|\alpha|^2+4\left[(1+2\nu)(1-\nu)-2\sqrt{\nu(1-\nu)}\right]|\beta|^2\right].
\label{Eq:Soundness}
\end{equation}
The analogous soundness for the $M=2$ scheme with experimental imperfections reads:
\begin{equation}
s_2^{exp} =1 - \exp\left[-\eta\left(\nu - \frac{1}{2}\right)|\alpha-\beta|^2 -\eta\left(1 - \nu + \sqrt{\nu(1 - \nu)}\right)|\alpha|^2-\eta\left(1 - \nu - \sqrt{\nu(1 - \nu)}\right)|\beta|^2\right].
\end{equation}
\end{widetext}
In the absence of any experimental imperfections, it is straightforward to see that $s_4$ is always greater than $s_2$. We show that, even with experimental imperfections, $s_4^{exp} \ge s_2^{exp}$ for any visibility factor $\nu$ and any quantum efficiency $\eta$ (see Appendix~\ref{app:s2vss4}). Thus the $M = 4$ scheme outperforms the $M = 2$ scheme in soundness. We note the similar gain in the soundness is expected when comparing $M = 2$ scheme with the scheme involving any arbitrary M value. For simplicity, we have shown the comparison for $M=2$ and $M=4$ schemes.


\section{State comparison for an untrusted source} \label{sec:genCV}
 
\noindent We now consider an adversarial scenario where the unknown state in the input register is not restricted to being a coherent state, but can be any generic (mixed)  quantum state produced by some untrusted party, while the states in the program register are coherent states obtained from a trusted source. The task is then to check whether the states in the input and program registers are equal. We first analyse this state comparison task under one-sided error when the program register contains a single coherent state $\ket{\beta}$. Subsequently, we generalise the state comparison procedure by allowing the multiple copies of $\ket{\beta}$ in the program register.  We note that in both settings, we receive only a single unknown state from the untrusted source.  We conclude this section by proving that our scheme is optimal for state comparison even when we do not make any assumption whatsoever about the state in the input register. 

\subsection{State comparison with a single copy of the test and program register states}

\noindent We consider the scenario where the input register state is a generic quantum state $\tau$ and the program register state is a coherent state $\ket{\beta}$. Any single-mode state $\tau$ can be expressed in the Fock basis as
\begin{equation}
\tau = \sum_{k,l\ge0}{\tau_{kl}\ket k\bra l},
\label{eq:tau1}
\end{equation}
with the normalization condition $\sum_{n\ge0}{\tau_{nn}}=1$, coming from from Tr$(\tau) = 1$.

Let us look at the completeness and soundness arguments again:\\

\noindent\textbf{Completeness}:
 If the states are $\tau$ and $\ket{\beta}$ are the same, then their trace distance is $0$. Thus, the probability of having the detection event 1 is zero.
 This ensures perfect completeness again, i.e., $c_2 = 1$.\\
 
\noindent\textbf{Soundness}:
\textit{Trace distance.} Suppose $\|\tau - \ket{\beta}\bra{\beta}\|_{\text{tr}} \ge \epsilon$. This implies
\begin{equation}
\sqrt{1-\bra{\beta}\tau\ket{\beta}} \geq \epsilon,
\label{eq:trace1}
\end{equation}
so $\bra{\beta}\tau\ket{\beta}\leq 1 - \epsilon^2$. We show in Appendix~\ref{app:newproofsoundness2} that the probability of obtaining a click in the detector $D_0$ after the interaction of the states $\tau$ and $\ket{\beta}$ in the balanced beam splitter is tightly lower bounded as
\begin{equation}
\begin{aligned}
\mathbb P_{D_1} &\geq\frac12(1-\bra{\beta}\tau\ket{\beta})\\
&\geq\frac{\epsilon^2}2.
\end{aligned}
\label{eq:clickd2}
\end{equation}
Thus the soundness in this case is $s_2 \geq  \frac{\epsilon^2}{2}$.

\subsection{Generalised single run state comparison}

\noindent The generalised single run state comparison scheme is run on a single unknown state $\tau$ (Eq.~(\ref{eq:tau1})) in the input register and $M-1$ coherent states $\ket{\beta}$ in the program register. 
Here we analyse the completeness and soundness of the \emph{amplifier scheme} used for comparison as described in section~\ref{ampscheme}.\\

\noindent \textbf{Completeness}: If the states $\tau$ and $\ket{\beta}$ are the same, then their trace distance is 0, and the probability of a click in at least one of the detectors in output modes $2^k$ is 0. Thus we obtain perfect completeness $c_{M} = 1$. 
\\\\
\noindent \textbf{Soundness}: Suppose $\|\tau - \ket{\beta}\bra{\beta}\|_{\text{tr}} \ge \epsilon$, so $\bra{\beta}\tau\ket{\beta}\leq 1 - \epsilon^2$. We show in Appendix~\ref{app:newproofsoundnessM} that the probability that at least one of the detectors in output modes $2^k$ clicks after the interaction of the states $\tau$ and $\ket{\beta}$ in the balanced beam splitter is tightly lower bounded as
\begin{equation}
\begin{aligned}
\mathbb P_{D} &\geq\left(1 - \frac{1}{M}\right)(1-\bra{\beta}\tau\ket{\beta})\\
&\geq\left(1 - \frac{1}{M}\right)\epsilon^2.
\end{aligned}
\label{eq:clickdM}
\end{equation}
Thus the soundness in this case is $s_M \geq \left(1 - \frac{1}{M}\right)\epsilon^2 > s_2$ for $M > 2$. The bound improves by adding more copies of the program register states, as can be expected.

\subsection{Test optimality}

\noindent The proof of optimality derived in section~\ref{app:opti} holds even when the input register state is a generic mixed state, as long as the program register states are coherent states. Indeed, the optimal POVM for state comparison, when one has a single copy of input register state and $M-1$ copies of the program register states, is derived assuming it satisfies the completeness relation in Eq.~(\ref{cond1}), which is also the case here. This implies that the optimal POVM when the tested state is generic, while the program register states are coherent states, is the same as the one constructed in section~\ref{app:opti}. This proves the optimality of our proposed projective schemes, in this generalised setting.


\section{Improved Quantum fingerprinting} \label{fingerprint}

\noindent As a concrete example of an application of our generalised state comparison scheme, we consider the improvement in performance of a specific quantum communication protocol: the quantum fingerprinting protocol for estimating the Euclidean distance of two real vectors within a constant factor. Our model of study is the simultaneous message passing model of communication complexity \cite{kumar2017efficient}.

The communication task is as follows. Two parties Alice and Bob receive data sets $x$ and $y$ respectively, which are unit vectors in $\mathbb R^n$. They are interested to check the similarity of their data sets, through a Referee, by estimating the (square of the) Euclidean distance of their vectors, $||x-y||_2^2 = \sum_{j=1}^n (x_j-y_j)^2$ within some multiplicative constant $\epsilon$ with a probability at least $1 - \delta$.

A trivial solution to this problem would be Alice and Bob transmitting the strings $x$ and $y$ respectively to the Referee. This however is a non-optimal protocol in terms of communication resources (number of bits sent to Referee) when the task is only to approximate the Euclidean distance. As we show, the task can be solved with much lower communication resources when Alice and Bob send the fingerprints of their data sets, which would typically be of much shorter length while still allowing the Referee to estimate the Euclidean distance within some constant.  When we restrict the model to Alice and Bob sharing no randomness, the classical fingerprint size necessary to solve this problem is $\Omega(\sqrt{n})$ \cite{ambainis1996communication,babai1997randomized,newman1991private,newman1996public}.

Motivated by the original quantum fingerprinting protocol of Buhrman et al.~\cite{buhrman2001quantum} to check for the equality of two $n$-bit strings, Kumar et al.~\cite{kumar2017efficient} proposed a coherent state quantum fingerprinting protocol to estimate the Euclidean distance with $\mathcal{O}(\log n)$ qubits which is asymptotically exponentially shorter in size compared to the classical fingerprints. We note that a similar improvement in resources for the approximately checking the equality of two bit strings problem was proposed by \cite{arrazola2014quantum} in the coherent state framework and subsequently demonstrated in \cite{xu2015experimental, guan2016observation}.

\subsection{Quantum fingerprinting protocol to approximate the Euclidean distance}

\begin{figure}
\vspace{-0.1cm}
\includegraphics[scale=0.53]{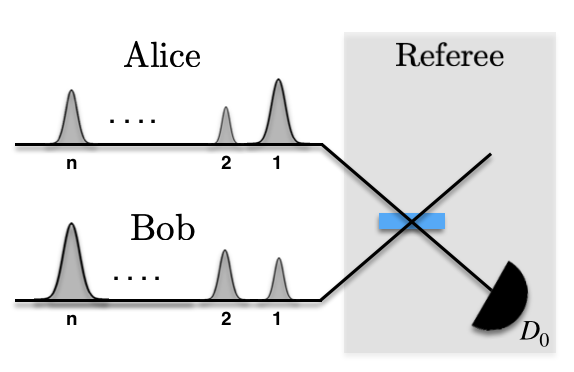}
\centering
\vspace{-0.4cm}
\caption{Alice and Bob prepare coherent state fingerprints, $\ket{1_x}$ and $\ket{1_y}$, as a sequence of coherent pulses in $n$ modes, with the $j^{th}$ mode amplitude encoding the information of $j^{th}$ component of the vector. This is then interfered sequentially in the balanced beam splitter and the results are analysed in single-photon threshold detector $D_0$.}
\label{fig:cs}
\end{figure}

\noindent Here we review the coherent state fingerprinting protocol proposed by \cite{kumar2017efficient} to approximate the Euclidean distance.  Alice and Bob prepare quantum fingerprints of their data sets, which are a sequence of coherent pulses in $n$ modes, and send these to the Referee. Alice (similarly Bob) prepares her state $\ket{1_{x}}$ by applying the displacement operator $\hat{D}_{x}(1) = \exp( \hat{a}_{x}^{\dagger} -  \hat{a}_{x})$ to the vacuum state, where $\hat{a}_{x} = \sum_{j=1}^{n}x_j\hat{b}_j$ is the superposed annihilation operator
~\cite{arrazola2014quantum}, and $\hat{b}_j$ is the photon annihilation operator of the $j^{th}$ mode.  The coherent state fingerprint of Alice is then,
\begin{equation}
\ket{1_{x}} = \hat{D}_{x}(1)\ket{0} = \otimes_{j=1}^{n}\ket{x_j}_j,
\end{equation}
where $\ket{x_j}_j$ is a coherent state of amplitude $x_j $ occupying the $j^{th}$ mode. The mean photon number for the state $\ket{1_{x}}$ is given by $\mu = \sum_{j} |x_j|^2 = 1$, independent of the input size. This scheme is illustrated in the Fig.~\ref{fig:cs}.

The Referee interacts the coherent fingerprints from Alice and Bob sequentially using a balanced beam splitter and observes the single-photon clicks in a threshold detector $D_0$. The input of the interaction is
\begin{equation}
\ket{1_x} \otimes \ket{1_y} = \bigotimes_{j=1}^{n} \ket{x_j}_j \otimes \ket{y_j}_j.
\end{equation}
The output in the mode corresponding to the detector $D_0$ at the $j^{th}$ pulse is $\ket{\frac{x_j - y_j}{\sqrt{2}}}_j$. Let $Z_j$ be the binary random variable which equals 1 with the probability that the detector $D_1$ clicks at the $j^{th}$ pulse, namely $p_j=1 - \exp(-\frac{(x_j-y_j)^2}{2})\approx \frac{(x_j-y_j)^2}{2}$. The latter approximation holds true since $x$ and $y$ are unit vectors in $\mathbb R^n$ and for large $n$ the terms $(x_j-y_j)^2$ are typically of the order of $1/n$. The Euclidean distance ($\tilde{E}$) is equal to
\begin{equation}\label{perfect}
\tilde{E} = 2\cdot\mathbb{E}\left[\sum_{j=1}^n Z_j\right].
\end{equation}
The Chernoff-Hoeffding inequality \cite{upfal2005probability} can then be used to estimate $\tilde{E}$ with $\sum_{j=1}^n Z_j$ within a mutiplicative precision $\epsilon>0$:
\begin{equation}
\mathbb{P}\left[\left|\sum_{i=1}^{n}Z_j - \frac{\tilde{E}}{2}\right| \geq \epsilon\frac{\tilde{E}}{2}\right]\leq 2\exp\left(-\frac{\epsilon^2\tilde{E}}{6}\right)
\label{chernoff}
\end{equation}
Eq.~(\ref{chernoff}) provides the Referee with the estimation of the Euclidean distance between two unit vectors $x$ and $y$ within a desired multiplicative precision $\epsilon$, with a success probability at least $1 - 2\delta$, with $\delta = \exp(-\frac{\epsilon^2\tilde{E}}{6})$ in a single run.
\begin{figure*}[t]
\includegraphics[scale=0.5]{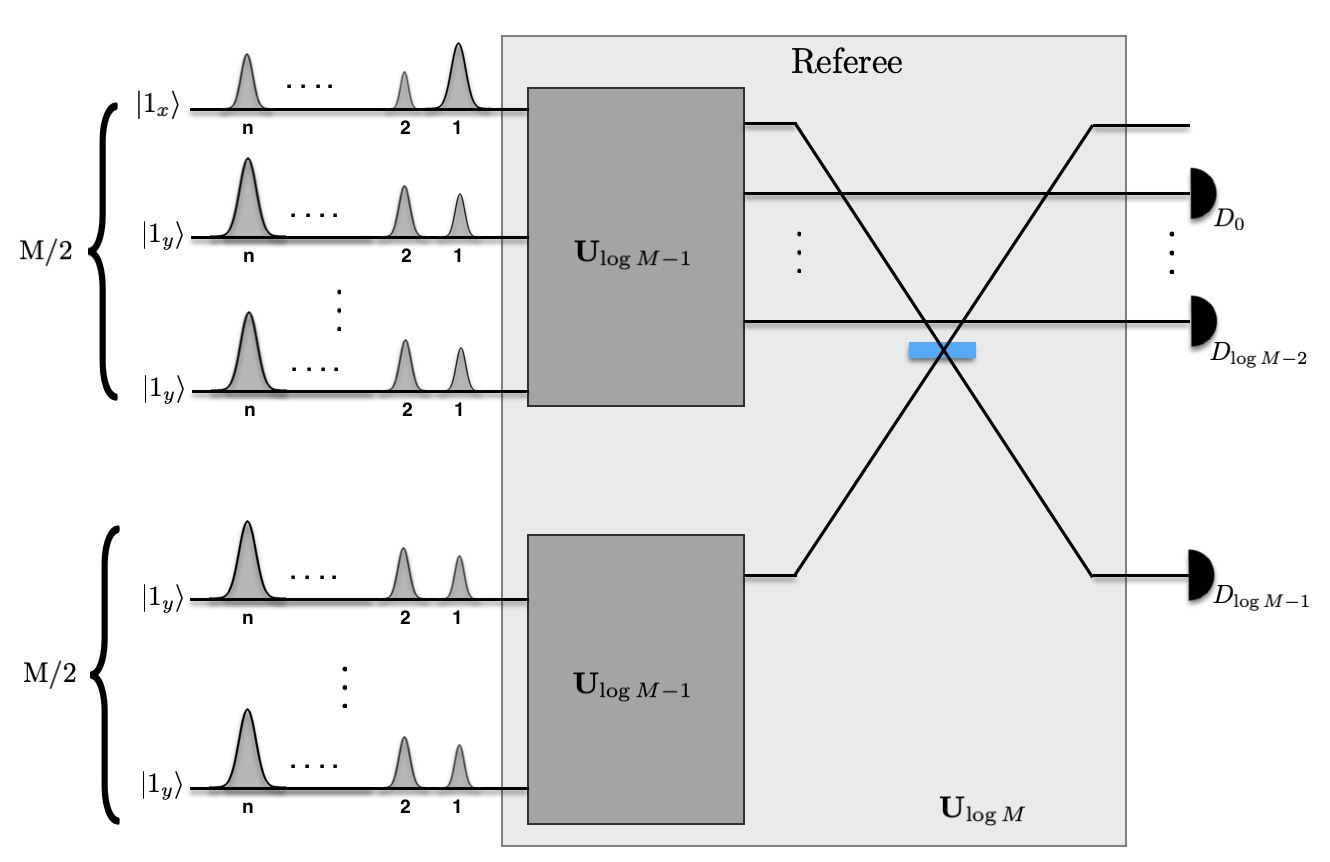}
\centering
\caption{Improved quantum fingerprinting protocol where Alice sends a single copy of $\ket{1}_x$ to the Referee while Bob sends $M -1$ copies of the fingerprint $\ket{1_y}$ to the Referee. Here the Referee employs the \emph{amplifier scheme} with $M-1$ beam splitters and $\log M$ threshold detectors $\{D_{0},D_{1},\dots, D_{\log M -1 }\}$.}
\label{fig:Geneuclid}
\end{figure*}

For the coherent state fingerprint with an average photon number $\mu = 1$ across $n$ time modes, the fingerprint size (transmitted information) is $\mathcal{O}(\log n)$. Thus the estimation of the Euclidean distance within $(\epsilon, \delta)$, requires Alice and Bob each to send exponentially smaller sized fingerprints to the Referee than the classical analogue of $\mathcal{O}(\sqrt{n})$.

\subsection{Improved quantum fingerprint protocol}

\noindent Here we consider the performance of our generalised state comparison method and demonstrate that in a single run, the Referee is able to estimate the Euclidean distance of two vectors $x,y \in \mathbb R^n$ with substantially better probability than the original fingerprinting protocol.

The setting is as follows: 
Similar to the protocol of \cite{kumar2017efficient}, Alice and Bob prepare their quantum fingerprints $\ket{1_x}$ and $\ket{1}_y$ respectively as sequence of coherent pulses in $n$ modes. Here, one of the parties, say Alice, sends a single copy of the fingerprint state to the Referee, while Bob sends multiple, $M-1$, copies of the fingerprint state to the Referee. Fig.~\ref{fig:Geneuclid} illustrates the setting where upon receiving the fingerprint states, the Referee applies the projective measurement scheme. Here we showcase the \emph{amplifier scheme}, but the same results are obtained using the \emph{Hadamard} and \emph{looped amplifier} schemes.

The referee applies the generalised amplifier interferometer $U_{\log M}$ of Eq.~(\ref{U}) to the incoming states. The clicks are then collected in the $\log M$ threshold detectors indexed as $D :\{D_{0},D_{1},\dots, D_{\log M-1}\}$. \par

Let $Z_{j,k}$ be the binary random variable that is 1 with the probability that the detector indexed $D_{k}$ clicks at the $j^{th}$ pulse, with $j \in \{1,\dots,n\}$ and $k \in \{0,\dots,\log M -1\}$, namely $p_{j,k} = 1 - \exp(-\frac{(x_j-y_j)^2}{2^{k+1}})\approx \frac{(x_j-y_j)^2}{2^{k+1}}$. Again the approximation holds true since $x$ and $y$ are unit vectors in $\mathbb R^n$ and for large $n$ the terms $(x_j - y_j)^2$ are typically of the order of $1/n$. The Euclidean distance ($\tilde{E}$) in this case is equal to,
\begin{equation}
\tilde{E} = \frac{1}{1 -\frac1M}\cdot\mathbb{E}\left[\sum_{j=1,k=0}^{n, \log M-1}Z_{j,k}\right].
\end{equation}
Once again, the Chernoff-Hoeffding inequality can be used to estimate $\tilde{E}$ with $\sum_{j=1,k=0}^{n, \log M-1}Z_{j,k}$ within a multiplicative precision $\epsilon>0$:
\begin{equation}
\ba
\mathbb{P}&\left[\left|\sum_{j=1,k=0}^{n, \log M-1}Z_{j,k} -\left(1 - \frac{1}{M}\right)\tilde{E}\right| \geq \epsilon\left(1 - \frac{1}{M}\right)\tilde{E} \right]\\
&\quad\quad\quad\quad\quad\quad\leq 2\exp\left(-\epsilon^2\left(1-\frac1M\right)^2 \frac{\tilde{E}}{3}\right)\\
&\quad\quad\quad\quad\quad\quad\le2\delta^{2\left(1-\frac1M\right)},
\ea
\label{chernoff2}
\end{equation}
where $2\delta$ is the failure probability in estimating the Euclidean distance in the original fingerprint protocol.

With the improved quantum fingerprinting protocol, we conclude with Eq.~(\ref{chernoff2}) that the Euclidean distance between two vectors $x$ and $y$ can be estimated within an additive factor $\epsilon$ with a probability $1-2\delta^{2(1 - \frac{1}{M})}$ which is better than the original protocol by a factor $2\left(1 - \frac{1}{M}\right) > 1$ for all $M >2$. For example, $M=4$ already provides a power $1.5$ improvement in the probability of successfully estimating the Euclidean distance.

We note that when Bob sends $M-1$ coherent state fingerprints to the Referee, this leads to the total transmitted information to be $M\times \mathcal{O}(\log n)$. As long as $M$ is independent of $n$, we still achieve an asymptotic exponential reduction in transmitted information compared to the classical analogue of $\mathcal{O}(\sqrt{n})$.

\section{Discussion}
\label{sec:concl}

\noindent We have presented an optimal programmable measurement scheme that projects the incoming single mode state in the input register onto a local coherent state in the program register. Our scheme is implemented using balanced beam splitters and single-photon threshold detectors. Our implementation of this interferometer is efficient, and threshold detectors with high efficiency and low dark counts are commercially available \cite{schweickert2018demand}.
Our most efficient scheme in terms of optical components, the \emph{looped amplifier scheme}, requires only a single beam splitter and a single threshold detector, together with an optical switch. 
We have further generalised our schemes to a cryptographic setting where the input register state is obtained from an untrusted source, and is no longer assumed to be a coherent state.

This universal implementation using coherent states as program states can act as a backbone in improving the performance of a range of quantum protocols in communication complexity \cite{buhrman2001quantum, de2004one}, cryptography and computational regimes \cite{ekert2002direct,mintert2005concurrence, walborn2006experimental, aaronson2008power,harrow2013testing, lloyd2013quantum}. For example, it has been shown that communication protocol using coherent state fingerprints provides an asymptotic exponential advantage in communication resources in estimating the Euclidean distance between two vectors compared to any classical protocol \cite{kumar2017efficient}. For this protocol, we show a further quadratic improvement in the probability of correctly estimating the Euclidean distance compared to the original protocol.
 
The general applicability of our optimal state comparison schemes makes it attractive to be readily implementable. Another feature that we bring is that the output quantum states are not completely destroyed after the performing the comparison test. Further, the output states carry the overlap information of the unknown quantum test. 


\section{Acknowledgements}

\noindent We thank Frederic Grosshans for useful discussions. 
N.K. and E.K. acknowledge the support of UK Engineering and Physical Sciences Research Council grant  EP/N003829/1. D.M. acknowledges the funding from the ANR through the ANR-17-CE24-0035 VanQuTe project. E.D. acknowledges financial support from the European Research Council Starting Grant QUSCO.

\section{Author Contributions}

N.K. and U.C. conceptualized the idea and wrote the proofs. 
All authors contributed in preparing the manuscript.  

\bibliographystyle{apsrev}
\bibliography{bibliography}


\newpage


\onecolumngrid
\appendix


\section{Computing the expression of the operator $\Delta_M$ in Fock basis}
\label{app:Delta}

\noindent Using the notations of the main text, and writing,
\be
\ket\alpha=e^{-\frac{|\alpha|^2}{2}}\sum_{n=0}^{+\infty}{\frac{\alpha^n}{\sqrt{n!}}\ket n},
\ee
we obtain
\begin{equation}
\begin{aligned}
\Delta_M&=\int{d^2\alpha\text{ } \exp[-M|\alpha|^2]\sum_{\substack{k_j,l_j=0\\\forall j\in [M]}}^{\infty}{\frac{\alpha^{\sum_j k_j}(\alpha^*)^{\sum_j l_j}}{\sqrt{k_1!\dots k_M!l_1!\dots l_M!}}\ket{k_1\dots k_M}\bra{l_1\dots l_M}}}\\
&=\sum_{\substack{k_j,l_j=0}}^{\infty}\frac{\ket{k_1\dots k_M}\bra{l_1\dots l_M}}
{\sqrt{k_1!\dots k_M!l_1!\dots l_M!}}\int_{r=0}^{\infty}{dr\text{ }\exp[{-Mr^2}]r^{1+\sum_j k_j+l_j}}\int_{\theta=0}^{2\pi}{d\theta\text{ }\exp[{i\theta\sum_j (k_j-l_j)}]}\\
&=\frac{\pi}{M}\sum_{\substack{k_j,l_j=0}}^{\infty}\frac{\delta_{\sum_j k_j,\sum_j l_j}}{M^{\frac{\sum_j k_j}{2}}M^{\frac{\sum_j l_j}{2}}}\sqrt{\frac{(\sum_j k_j)!(\sum_j l_j)!}{k_1!\dots k_M!l_1!\dots l_M!}}\ket{k_1\dots k_M}\bra{l_1\dots l_M}\\
&=\frac{\pi}{M}\sum_{N=0}^{\infty}{\sum_{\substack{\sum_j k_j=N\\\sum_j l_j=N}}{M^{-N}\sqrt{\frac{N!}{k_1!\dots k_M!}}\sqrt{\frac{N!}{l_1!\dots l_M!}}\ket{k_1\dots k_M}\bra{l_1\dots l_M}}}\\
&=\frac{\pi}{M}\sum_{N=0}^{\infty}{\ket{\chi_N^M}\bra{\chi_N^M}},
\end{aligned}
\end{equation}
where we have defined for all $N\ge0$,
\begin{equation}
\ket{\chi_N^M}=M^{-N/2}\sum_{\sum_j k_j=N}{\sqrt{\frac{N!}{k_1!\dots k_M!}}\ket{k_1\dots k_M}}.
\end{equation}
%


\section{Computing the soundness for $M=4$}
\label{app:s4}

\noindent When the states are different, the probability of obtaining the detection event $0$ (failure probability) is
\be
P_f(M=4)=1-s_4=\exp{\left(-\frac{\eta}{4}A\right)},
\ee
where
\be
\ba
A&=2\left|\sqrt\nu(\alpha-\beta)+\sqrt{1-\nu}(\alpha+\beta)\right|^2+\left|\alpha-\beta+2\sqrt{\nu(1-\nu)}(\alpha+\beta)\right|^2\\
&=(1+2\nu)|\alpha-\beta|^2+2(1+2\nu)(1-\nu)|\alpha+\beta|^2+8\sqrt{\nu(1-\nu)}(|\alpha|^2-|\beta|^2),
\ea
\ee
where we used $(\alpha-\beta)(\alpha+\beta)^*+(\alpha-\beta)^*(\alpha+\beta)=2|\alpha|^2-2|\beta|^2$. Using $|\alpha+\beta|^2=2|\alpha|^2+2|\beta|^2-|\alpha-\beta|^2$ we obtain
\be
A=(4\nu^2-1)|\alpha-\beta|^2+4\left[(1+2\nu)(1-\nu)+2\sqrt{\nu(1-\nu)}\right]|\alpha|^2+4\left[(1+2\nu)(1-\nu)-2\sqrt{\nu(1-\nu)}\right]|\beta|^2.
\ee
%


\section{Comparing the soundness for $M=2$ to the soundness for $M=4$}
\label{app:s2vss4}

\noindent We show in this section that $s_2^{exp}\le s_4^{exp}$ for all $\alpha,\beta$. We have
\begin{equation}
\begin{aligned}
s_2^{exp} ={} & 1 - \exp\bigg[-\eta\bigg(\nu - \frac{1}{2}\bigg)|\alpha-\beta|^2 -\eta\bigg(1 - \nu + \sqrt{\nu(1 - \nu)}\bigg)|\alpha|^2 \\
&\hspace{17mm} -\eta\bigg(1 - \nu - \sqrt{\nu(1 - \nu)}\bigg)|\beta|^2\bigg]\\
&\equiv1-\exp[-\eta A_2],
\end{aligned}
\end{equation}
and
\begin{equation}
\small
\begin{aligned}
s_4^{exp} ={} & 1 - \exp\bigg[-\eta\bigg(\nu^2-\frac{1}{4}\bigg)|\alpha-\beta|^2 \\
&\hspace{15mm} -\eta\bigg((1+2\nu)(1-\nu)+2\sqrt{\nu(1-\nu)}\bigg)|\alpha|^2 \\
&\hspace{15mm} -\eta\bigg((1+2\nu)(1-\nu)-2\sqrt{\nu(1-\nu)}\bigg)|\beta|^2\bigg]\\
&\equiv1-\exp[-\eta A_4].
\end{aligned}
\end{equation}
Since the function $x\mapsto 1-\e^{-x}$ is increasing, it is sufficient to show that $A_2\le A_4$ for all $\alpha,\beta$. Writing $\alpha=re^{i\phi}$ and $\beta=te^{i\psi}$, where $r,t\ge0$ and $\phi,\psi\in[0,2\pi]$, we obtain
\begin{equation}
\small
\begin{aligned}
A_4-A_2&=\bigg(\frac{1}{4}+\nu(1-\nu)+\sqrt{\nu(1-\nu)}\bigg)r^2 \\
&\hspace{15mm} + \bigg(\frac{1}{4}+\nu(1-\nu)-\sqrt{\nu(1-\nu)}\bigg)t^2\\
&\hspace{15mm} -2rt\bigg(\frac{1}{4}-\nu(1-\nu)\bigg)\cos(\phi-\psi).
\end{aligned}
\end{equation}
This last expression is a polynomial of degree $2$ in $r$, with a positive leading coefficient. Thus if its discriminant is negative, then the expression is always positive. The discriminant is
\be
\ba
\Delta&=4t^2\bigg[\bigg(\frac{1}{4}-\nu(1-\nu)\bigg)^2\cos(\phi-\psi)^2\\
&\hspace{15mm} -\bigg(\frac{1}{4}+\nu(1-\nu)+\sqrt{\nu(1-\nu)}\bigg)\bigg(\frac{1}{4}+\nu(1-\nu)-\sqrt{\nu(1-\nu)}\bigg)\bigg]\\
&\le-6t^2\nu(1-\nu)\\
&\le0,
\ea
\ee
where the third line is obtained by using $\cos(\phi-\psi)\le1$.
Hence for all experimental parameters within the error model we consider, we have $s_2^{exp}\le s_4^{exp}$.


\section{Soundness for a generic input state with $M=2$}
\label{app:newproofsoundness2}

\noindent By linearity of the probabilities, it is sufficient to prove Eq.~(\ref{eq:clickd2}) when $\tau$ is a pure state.

Let us write $\tau=\ket\psi\bra\psi$, where $\ket\psi=\sum_{n\ge0}{\psi_n\ket n}$, where $\sum_{n\ge0}{|\psi_n|^2}=1$. We first show Eq.~(\ref{eq:clickd2}) for $\beta=0$. In that case, the $2$-mode input state is
\begin{equation}
\begin{aligned}
\ket{\psi}_a\otimes\ket{0}_b&=\sum_{n\ge0}{\psi_n\ket{n}_a\otimes\ket{0}_b}\\
&=\sum_{n\ge0}{\frac{\psi_n}{\sqrt{n!}}(\hat a^\dag)^n\ket{0}_a\otimes\ket{0}_b},
\end{aligned}
\end{equation}
where $\hat a^\dag$ is the creation operator corresponding to the input mode of $\ket\psi$.
Writing $\hat c^\dag$ and $\hat d^\dag$ the creation operators of the output modes after the balanced beam splitter evolution $\hat H$, the output state is given by:
\begin{equation}
\begin{aligned}
\hat H\ket{\psi}_a\otimes\ket{0}_b&=\sum_{n\ge0}{\frac{\psi_n}{\sqrt{n!}}\left(\frac{\hat c^\dag+\hat d^\dag}{\sqrt2}\right)^n\ket{0}_c\otimes\ket{0}_d}\\
&=\sum_{n\ge0}\frac{\psi_n}{2^{n/2}\sqrt{n!}}\sum_{k=0}^n{\binom nk(\hat c^\dag)^k(\hat d^\dag)^{n-k}\ket{0}_c\otimes\ket{0}_d}\\
&=\sum_{n\ge0}\frac{\psi_n}{2^{n/2}}\sum_{k=0}^n{\sqrt{\binom nk}\ket{k}_c\otimes\ket{n-k}_d}.
\end{aligned}
\end{equation}
The probability that the detector $D_0$, corresponding to the output mode $d$, does not click is given by:
\be
\ba
\mathbb P_{\emptyset}&=\text{Tr}\left[(\mathbb I_c\otimes\ket0\bra0_d)\hat H\ket\psi\bra\psi_a\otimes\ket0\bra0_b\hat H^\dag\right]\\
&=\sum_{n,m\ge0}\frac{\psi_n\psi_m^*}{2^{(n+m)/2}}\sum_{k=0}^n\sum_{l=0}^m\sqrt{\binom nk\binom ml}\text{Tr}\left[(\mathbb I_c\otimes\ket0\bra0_d)(\ket{k}\bra l_c\otimes\ket{n-k}\bra{m-l}_d)\right]\\
&=\sum_{n\ge0}\frac{|\psi_n|^2}{2^n}\\
&\le|\psi_0|^2+\frac12\sum_{n\ge1}|\psi_n|^2\\
&=|\psi_0|^2+\frac12(1-|\psi_0|^2)\\
&=\frac12+\frac12|\psi_0|^2.
\ea
\ee
Note that this inequality is an equality whenever $\psi_n=0$ for $n>1$. Given that $\mathbb P_{D_1}=1-\mathbb P_{\emptyset}$, this yields
\be
\mathbb P_{D_0}\ge\frac12(1-|\!\braket{\psi|0}\!|^2),
\ee
which concludes the proof when $\beta=0$.

Now if $\beta\neq0$, we define
\be
\ket\phi:=\hat D^\dag(\beta)\ket\psi,
\ee
where $\hat D$ is a displacement operator, with $\ket\beta=D(\beta)\ket0$, so that $\ket\psi=D(\beta)\ket\phi$ and $|\!\braket{\psi|\beta}\!|^2=|\!\braket{\phi|0}\!|^2$. The input state is given by %
\be
\ket{\psi}_a\otimes\ket\beta_b=\hat D_a(\beta)\otimes\hat D_b(\beta)\ket{\phi}_a\otimes\ket{0}_b,
\ee
where the subscript indicates the modes onto which the displacement operator acts. The probability that the detector $D_0$ does not click after the beam splitter interaction is then given by:
\be
\ba
\mathbb P_{\emptyset}&=\text{Tr}\left[(\mathbb I_c\otimes\ket0\bra0_d)\hat H\ket\psi\bra\psi_a\otimes\ket0\bra0_b\hat H^\dag\right]\\
&=\text{Tr}\left[(\mathbb I_c\otimes\ket0\bra0_d)\hat H(\hat D_a(\beta)\otimes\hat D_b(\beta))\ket{\phi}\bra{\phi}_a\otimes\ket0\bra0_b(\hat D_a^\dag(\beta)\otimes\hat D_b^\dag(\beta))\hat H^\dag\right].
\ea
\ee
Now we have
\be
\hat H(\hat D_a(\beta)\otimes\hat D_b(\beta))=(\hat D_c(\sqrt2\beta)\otimes\mathbb I_d)\hat H.
\ee
Hence,
\be
\ba
\mathbb P_{\emptyset}&=\text{Tr}\left[(\mathbb I_c\otimes\ket0\bra0_d)(\hat D_c(\sqrt2\beta)\otimes\mathbb I_d)\hat H\ket{\phi}\bra{\phi}_a\otimes\ket0\bra0_b\hat H^\dag(\hat D_c^\dag(\sqrt2\beta)\otimes\mathbb I_d)\right]\\
&=\text{Tr}\left[(\mathbb I_c\otimes\ket0\bra0_d)\hat H\ket{\phi}\bra{\phi}_a\ket0\bra0_b\hat H^\dag\right],
\ea
\ee
and the previous proof for $\beta=0$ gives
\be
\mathbb P_{D_0}\ge\frac12(1-|\!\braket{\phi|0}\!|^2).
\ee
Finally, since $|\!\braket{\psi|\beta}\!|^2=|\!\braket{\phi|0}\!|^2$, we obtain
\be
\mathbb P_{D_0}\ge\frac12(1-|\!\braket{\psi|\beta}\!|^2),
\ee
and this inequality is an equality whenever $\braket{\phi|n}=0$ for $n>1$, with $\ket\phi=\hat D^\dag(\beta)\ket\psi$.


\section{Soundness for a generic input state for any $M$}
\label{app:newproofsoundnessM}

\noindent By linearity of the probabilities, it is sufficient to prove Eq.~(\ref{eq:clickdM}) when $\tau$ is a pure state.

Let us write $\tau=\ket\psi\bra\psi$, where $\ket\psi=\sum_{m\ge0}{\psi_m\ket m}$, where $\sum_{m\ge0}{|\psi_m|^2}=1$. We first show Eq.~(\ref{eq:clickdM}) for $\beta=0$. In that case, the $M$-mode input state is
\begin{equation}
\ket{\psi\,0\dots0}=\sum_{m\ge0}{\psi_m\ket{m0\dots0}}.
\label{Fockbasis}
\end{equation}
The probability that none of the detectors in output modes $2^k$ clicks (the modes being indexed from $0$ to $M-1$) after the amplifier interferometer evolution $\hat U_n$ is given by:
\be
\mathbb P_{\emptyset}=\text{Tr}\left[\left(\bigotimes_{i=0}^{M-1}\hat E_i\right)\hat U_n\ket{\psi\,0\dots0}\bra{\psi\,0\dots0}\hat U_n^\dag\right],
\ee
where $E_i=\ket0\!\bra0$ if $i=2^k$ and $E_i=\mathbb I$ otherwise. From section~\ref{app:opti}, since the amplifier scheme $U_n$ reproduces the statistics of the Hadamard scheme $H_n$, this probability can be written as
\be
\mathbb P_{\emptyset}=\text{Tr}\left[\Pi^{opt}_0\ket{\psi\,0\dots0}\bra{\psi\,0\dots0}\right],
\label{probaemptyinter}
\ee
where
\be
\Pi^{opt}_0=\sum_{N=0}^{\infty}{\ket{\chi_N^M}\bra{\chi_N^M}},
\ee
with
\be
\ket{\chi_N^M}:=M^{-N/2}\sum_{\sum_j k_j=N}{\sqrt{\frac{N!}{k_1!\dots k_M!}}\ket{k_1\dots k_M}}.
\ee
Expanding Eq.~(\ref{probaemptyinter}) in Fock basis using Eq.~(\ref{Fockbasis}) yields
\be
\ba
\mathbb P_{\emptyset}&=\sum_{N=0}^{\infty}{\sum_{\sum_i k_i=N}{\sum_{\sum_j l_j=N}{\sum_{p,q\ge0}{M^{-N}\sqrt{\frac{N!}{k_1!\dots k_M!}}\sqrt{\frac{N!}{l_1!\dots l_M!}}\psi_p\psi_q^*
\text{Tr}\left[(\ket{k_1\dots k_M}\bra{l_1\dots l_M})(\ket{p\,0\dots0}\bra{q\,0\dots0})\right]}}}}\\
&=\sum_{p\ge0}{\frac{|\psi_p|^2}{M^p}}\\
&\le|\psi_0|^2+\frac1M\sum_{p\ge1}{|\psi_p|^2}\\
&=|\psi_0|^2+\frac1M(1-|\psi_0|^2)\\
&=\frac1M+\left(1-\frac1M\right)|\psi_0|^2.
\ea
\ee
Note that this inequality is an equality whenever $\psi_n=0$ for $n>1$. Given that $\mathbb P_{D}=1-\mathbb P_{\emptyset}$, this gives
\be
\mathbb P_{D}\ge\left(1-\frac1M\right)(1-|\!\braket{\psi|0}\!|^2),
\ee
which concludes the proof when $\beta=0$.

Now if $\beta\neq0$, we define
\be
\ket\phi:=\hat D^\dag(\beta)\ket\psi,
\ee
as in the previous section, where $\hat D$ is a displacement operator, with $\ket\beta=D(\beta)\ket0$, so that $\ket\psi=D(\beta)\ket\phi$ and $|\!\braket{\psi|\beta}\!|^2=|\!\braket{\phi|0}\!|^2$. The input state is given by 
\be
\ket{\psi\,\beta\dots\beta}=\hat D(\beta)^{\otimes M}\ket{\phi\,0\dots0}.
\ee
The probability that none of the detectors in output modes $2^k$ clicks is given by:
\be
\ba
\mathbb P_{\emptyset}&=\text{Tr}\left[\left(\bigotimes_{i=0}^{M-1}\hat E_i\right)\hat U_n\ket{\psi\,\beta\dots\beta}\bra{\psi\,\beta\dots\beta}\hat U_n^\dag\right]\\
&=\text{Tr}\left[\left(\bigotimes_{i=0}^{M-1}\hat E_i\right)\hat U_n\hat D(\beta)^{\otimes M}\ket{\phi\,0\dots0}\bra{\phi\,0\dots0}\hat D^\dag(\beta)^{\otimes M}\hat U_n^\dag\right],
\ea
\ee
where $\hat E_i=\ket0\bra0$ if $i=2^k$ and $\hat E_i=\mathbb I$ otherwise. 
We have
\be
\hat U_n\hat D(\beta)^{\otimes M}=\left(\bigotimes_{i=0}^{M-1}{\hat D(\delta_i)}\right)\hat U_n,
\ee
where $(\delta_0,\dots,\delta_{M-1})^T=U_n(\beta,\dots,\beta)^T$. In particular, $\delta_i=0$ if $i=2^k$, so for all $i=0,\dots,M-1$,
\be
\hat D^\dag(\delta_i)\hat E_i\hat D(\delta_i)=\hat E_i.
\ee
Hence,
\be
\ba
\mathbb P_{\emptyset}&=\text{Tr}\left[\left(\bigotimes_{i=0}^{M-1}\hat E_i\right)\left(\bigotimes_{i=0}^{M-1}{\hat D(\delta_i)}\right)\hat U_n\ket{\phi\,0\dots0}\bra{\phi\,0\dots0}\hat U_n^\dag\left(\bigotimes_{i=0}^{M-1}{\hat D^\dag(\delta_i)}\right)\right]\\
&=\text{Tr}\left[\left(\bigotimes_{i=0}^{M-1}\hat D^\dag(\delta_i)\hat E_i\hat D(\delta_i)\right)\hat U_n\ket{\phi\,0\dots0}\bra{\phi\,0\dots0}\hat U_n^\dag\right]\\
&=\text{Tr}\left[\left(\bigotimes_{i=0}^{M-1}\hat E_i\right)\hat U_n\ket{\phi\,0\dots0}\bra{\phi\,0\dots0}\hat U_n^\dag\right].
\ea
\ee
The previous proof for $\beta=0$ gives
\be
\mathbb P_D\ge\left(1-\frac1M\right)(1-|\!\braket{\phi|0}\!|^2).
\ee
Finally, since $|\!\braket{\psi|\beta}\!|^2=|\!\braket{\phi|0}\!|^2$, we obtain
\be
\mathbb P_{D}\ge\left(1-\frac1M\right)(1-|\!\braket{\psi|\beta}\!|^2),
\ee
and this inequality is an equality whenever $\braket{\phi|n}=0$ for $n>1$, with $\ket\phi=\hat D^\dag(\beta)\ket\psi$.


\end{document}